\documentclass[a4paper,12pt]{article}

\usepackage[english]{babel}
\usepackage[T1]{fontenc}
\usepackage[utf8]{inputenc}
\usepackage[dvipsnames]{xcolor}

\usepackage{booktabs} 

\usepackage{amsmath, amsthm, amssymb, amsfonts}
\usepackage{mathptmx}
\usepackage{palatino}

\usepackage{graphicx}
\usepackage[sort&compress]{natbib}
\usepackage{arydshln}

\newcommand{\be}{\begin{equation}}
\newcommand{\ee}{\end{equation}}
\newcommand{\ban}{\begin{eqnarray*}}
\newcommand{\ean}{\end{eqnarray*}}
\def\is{\!\!\!\! & = & \!\!\!\!}

\def\half{\frac{1}{2}}

\newcommand{\ds}{\displaystyle}
\newcommand{\bea}{\begin{eqnarray}}
\newcommand{\eea}{\end{eqnarray}}
\newcommand{\bean}{\begin{eqnarray*}}
\newcommand{\eean}{\end{eqnarray*}}

\newtheorem{prop}{Proposition}[section]

\newcommand{\wtil}{\widetilde}

\newcommand{\e}{{\rm e}}
\newcommand{\dr}{{\mathrm d}}

\newcommand{\Q}{\mathbb{Q}}
\newcommand{\pp}{\mathbb{P}}

\usepackage[]{hyperref}

\hypersetup{
     bookmarks=false,        
     unicode=false,          
     pdftoolbar=true,        
     pdfmenubar=true,        
     pdffitwindow=false,     
     pdfstartview={FitH},    
     pdftitle={My title},    
     pdfauthor={},           
     pdfsubject={Subject},   
     pdfcreator={},          
     pdfproducer={},         
     pdfkeywords={Liquidity Risk}, 
     pdfnewwindow=true,      
     colorlinks=true,        
     linkcolor=red,        
     citecolor=blue,        
     filecolor=black,      
     urlcolor=magenta          
}

\begin{document}
\renewcommand\theequation{\thesection.\arabic{equation}}
\renewcommand{\thefootnote}{\fnsymbol{footnote}}
\begin{center}
{\LARGE Market Price of Trading Liquidity Risk \\ and 
\\[1em]
Market Depth
}
\\[0.5cm]

\vspace{1cm}

\newcommand*{\Authoremail}{christophert@smu.edu.sg}

\begin{center}
{\Large Masaaki Kijima and Christopher Ting} \\
School of Informatics and Data Science, 
Hiroshima University 
\end{center}

\vspace{1cm}

{\bf Abstract}
\end{center}

\noindent
Price impact of a trade is an important element
in pre-trade and post-trade analyses.
We introduce a framework to analyze 
the market price of liquidity risk,
which allows us to derive an inhomogeneous Bernoulli ordinary differential equation.
We obtain two closed form solutions,
one of which reproduces the linear function of the order flow in 
\cite{Kyle:1985} for informed traders.
However, when traders are not as asymmetrically informed, 
an S-shape function of the order flow is obtained.
We perform an empirical intra-day analysis on Nikkei futures
to quantify the price impact of order flow and compare our results with
industry's heuristic price impact functions.
Our model of order flow yields a rich framework for not only to estimate
the liquidity risk parameters,
but also to provide a plausible cause of 
why volatility and correlation are stochastic in nature.
Finally, we find that the market depth 
encapsulates the market price of liquidity risk.

\vskip10em

\newpage
\renewcommand{\thefootnote}{\arabic{footnote}}
\setcounter{footnote}{0}

\section{Introduction} \label{Introduction}
Whatever be the information content, it is no secret that transactions
induce the market to move in the direction that reflects the price
pressure over a time horizon.
In \cite{Perold:1988}, an institutional investor seeking to acquire
or liquidate a substantial amount of a security is likely
to cause the market price to deviate from the reference price\footnote{
The reference price is also known as the decision price 
or the arrival price.},
which is the market price prevailing at the time an investment decision
has just been made.
This deviation from the reference price---a
component cost in the framework of \cite{Perold:1988}---is a reaction 
of the market to the institutional investor's imminent need to trade.

Another cost of trading is known as the price impact cost or the market impact cost.
It is the compensation the liquidity providers receive for rendering
the service (of accommodation of mismatched order flow),
and the market impact cost paid by liquidity demanders 
(\cite{TF:1999}). 
According to \cite{Bouchaud:2010},
the price impact is the result of the correlation between 
an incoming market order and the subsequent price change,
in such a manner that the second buy
trade is on average more expensive than the first
because of its impact on the market price (and vice versa for sells).
In other words, price impact
is a post-trade explanation for the potential price disparity
when a market order or marketable limit order is executed.

Given the risk of price uncertainty and disparity from the target price,
what then is the market price of liquidity risk?
More importantly, what is the price impact function 
that captures the price disparity or slippage?
This paper attempts to provide some answers to these questions
by introducing a risk-neutral measure that is consistent
with the reality of no risk-free arbitrary opportunity.
We then propose the market price of liquidity risk in a form 
similar to the market price of (market) risk.
This idea forms the basis to obtain an analytical function
for measuring the price impact.
In other words, our analysis of the market price of liquidity risk
leads us to the derivations of the price impact functions.

When the electronic market is open for trading,
buy and sell orders are flowing into the exchange matching engine for execution.
Aggregating over a trading period, the order flow
(buyer-initiated trades less seller-initiated trades)
provides an ex post measure of net demand for the asset,
which translates into a market price.
This relationship between the order flow and price
is used by \cite{EL:2002} to show that order flow
is a superior candidate for modeling the variation
in foreign exchange rate compared to 
arguments along the lines of macroeconomics.

\clearpage
Even before trades occur, market participants are mindful that their
market orders may potentially impact prices; traders
will take into account the market impact of their  trades.
In other words, they know that there is going to be a disparity between
the price $P_t$ at which their market order $X_t$ is executed 
and the price $S_t$ at which they wish their orders can be executed immediately. 
More often than not, $P_t$ is disadvantageous than $S_t$ to the traders.
For that matter, price impact is created whenever $X_t \ne 0$
for all traders who consume liquidity.
Moreover, \cite{LP:2008} show that for major currency pairs,
using data sampled at the one-minute frequency, 
publicly announced macroeconomic information not only causes exchange rates to move, 
but also causes order flow to significantly change in a direction
consistent with the exchange rate movement.

The interest in price impact and order flow
spans several inter-related subjects of transaction costs,
optimal trading strategies (\cite{AC:2000} and \cite{Almgren:2012}), 
liquidity risk (\cite{AP:2005} and \cite{Amihud:2002}), 
price manipulation (\cite{GSS:2012}).
portfolio performance (\cite{DLLW:2016}),
payoff replication (\cite{LY:2005}),
as well as dynamic hedging (\cite{KFV:2016}).
Obviously, there are voluminous papers on these subjects,
and the list is by no means exhaustive.
This paper adds to this rich literature
by deriving from the market price of liquidity risk a functional form for
the price impact as an analytical solution.
Empirically, it is well known that
when the size of order flow is small,
the price impact is approximately linear.
But when the trade size is very large, 
nonlinear effects become pronounced:
concave when $X_t$ is positive and convex 
when $X_t$ is negative\footnote{See \cite{BFF:2009} 
and \cite{Bouchaud:2010} for a review.}. 

The formalism of this paper is similar to \cite{CJP:2004} and
\cite{HT:2008}, among others, in that traded prices are dependent on
the order size and another price referred to as the marginal price
or the true price.
The true price, in particular, originally appears in \cite{Roll:1984}.
In these papers, the true price is unobservable,
as it is the price corresponding 
to the absence of trades, i.e., zero order flow.
Furthermore, the literature of market microstructure is replete with
the assumption that the true price is a geometric Brownian motion.

This paper further postulates that order flows are mean reverting.
A standard transformation in stochastic calculus allows us 
to define the market price of liquidity.
In making the true price process to be under the risk-neutral measure
so as to reflect no risk-free arbitrage opportunity, 
we introduce the market price of (market) risk and
also the market price of liquidity risk.
We solve the Bernoulli differential equation and 
obtain a closed-form solution for the price impact function
and the market price of liquidity risk.
Both functions are dependent on the order flow.
These results lead us to a natural measure of liquidity risk,
much like volatility is a measure of market (price) risk.
It turns out that the parameters of the market price of liquidity risk
are encapsulated in a measure that can be interpreted
as the market depth of the electronic market.
The larger or deeper the market depth is, the market can be said to be more liquid.
 
Our model of price impact function is different from  
but complementary to the existing papers on price impact.
For example, \cite{WCBH:2015} empirically 
find that price impact exhibits a sub-linear power-law scaling
of  daily-normalized volume.
In \cite{BFF:2009}, the theory of long-term resilience is put forth
to model price impacts over different time lags.
\cite{Gerig:2007} also develops a theory of price impact, 
with a focus on hidden order.
He concludes that the impact of a hidden order is a concave function of its
total volume, specifically, a logarithm function.


It is well beyond the scope of this paper 
to conduct an exhaustive review of closely related research.
Nevertheless, the two above-mentioned theoretical works and other related papers
converge to the general consensus that the price impact of a trade
is a concave function of absolute volume normalized with certain heuristic schemes.
By contrast, our price impact function is an S-shape function of order flow.
Moreover, the other main difference is that our model 
is based on the market price of liquidity risk 
through the application of It\^{o}'s calculus.

Price impact is a major concern
for portfolio managers when they are about 
to reconstitute or re-balance their portfolios.
Many prime broker-dealers have their own proprietary methods 
to measure the price impact.
According to \cite{Ferraris:2008},
price impact models used in the industry include the following:
\begin{enumerate}  
\item BARRA model (\cite{TF:1999})
\[
\alpha\sqrt{V}
\]

\item Bloomberg's model
\[
\half \frac{\hbox{bid-ask spread}}{\hbox{price}}
+ \sqrt{\frac{\sigma^2/3}{250}} \sqrt{\frac{V}{0.3\times\hbox{EDV}}}
\]

\item JP Morgan's model 
\[
95\% \times \frac{1.4}{\hbox{EPV} \times \sqrt{\hbox{EDV}}} \sigma^2 V^{3/2}
+ 5\% \times \frac{0.187}{\sqrt{\hbox{EDV}}} \sigma^2 \sqrt{V} 
\]
\end{enumerate}
In these phenomenological models, $\alpha$ is the parameter to be estimated,
$\sigma$ is the volatility, and $V$ the volume.
Respectively,  EDV and EPV denote the expected daily volume 
and the expected period volume.

The first terms in the models of Bloomberg and JP Morgan
correspond to the temporary price impact,
and the second term is meant to estimate the permanent impact
(see \cite{Ferraris:2008}).
Temporary price impacts include the bid-ask spread.
They affect prices temporarily by disturbing the supply and demand equilibrium.
These transient disturbances are typically immediate at the point of transaction.
On the other hand, permanent price impacts are caused
by material information that will prompt investors 
to urgently update their belief about the value of the asset being traded.
They have a relatively long lasting effect on the subsequent prices.
Our price impact models capture mainly the temporary impact,
and indirectly the permanent impact through the correlation with the order flows.

The industry models, 
while consistent with the insights gleaned from 
the market microstructure literature, are heuristic in nature.
Generally, the price impact is assumed to be a square root of the trade size.
For example, \cite{Hasbrouck:2004} uses the square root of volume
to  measure the price impact of futures traded in the CME pit.
By contrast, our price impact function is derived from
the framework of market price of liquidity risk 
through the application of stochastic calculus.
It is based on the market reality of
no risk-free arbitrage opportunity.
Our model produces a linear price impact function 
of \cite{Kyle:1985} as a special case.
In the more general case, we find that the price impact function
is S-shape for all possible values of $X_t$.
Empirically, we find that our model provides a better fit to the same data.
Our price impact function may be a better model
to estimate the potential price impact in pre-trade and post-trade analyses.

Furthermore, by adding a mean-reverting model of order flow $X_t$
to the geometric Brownian motion of the true price $S_t$,
our modeling study shows that the randomness in $X_t$ causes
the volatility of simple returns computed from trade prices to become stochastic.
Likewise, the correlation between the simple return 
and the order flow has also become stochastic.

The paper will be presented as follows.
Section \ref{sect:Model Setup and Motivation}
provides the basic assumptions 
that parallel those of \cite{CJP:2004}. 
A key novel idea of our paper is to model the order flow dynamics
as a mean-reverting process.
Section \ref{Gen Analysis} introduces two market prices of risks,
leading to the inhomogeneous Bernoulli differential equation. 
In Section \ref{Analysis Leading to Price Impact Functions},
we consider two special cases and
derive the price impact functions in closed form.
We provide empirical evidence in Section \ref{Empirical Study} 
that demonstrates that our S-shape price impact function
is better than industry's heuristic functions.
Section \ref{Conclusions} concludes. 

\section{Model Setup and Motivation} \label{sect:Model Setup and Motivation}
\setcounter{equation}{0}
As defined by \cite{Lyons:2001}, 
order flow is transaction volume that is \emph{signed}.
A trade at the ask price is said to be buyer-initiated (positive)
and that at the bid price, seller-initiated (negative).
Market orders and marketable limit orders are means by which
transactions are initiated and executed.
With immediacy, these trades consume liquidity provided by dealers and market makers.
On electronic exchanges, 
liquidity is said to be provided by traders who submit limit orders,
which are displayed on an electronic limit order book.

Order flow has many nuances;
we define it for our model to avoid misunderstanding as follows.
Over a specified time interval, say 1 minute,
the order flow is computed by observed trades initiated by buyers and sellers.
Specifically, the order flow $X_t$ is written as
\be
X_t = X_t^+ - X_t^-,
\ee
where, over a time interval, 
$X_t^+$ is the aggregate volume of buyer-initiated trades
and $X_t^-$ is the aggregate trade size of seller-initiated trades.
In essence, $X_t$ reflects the net imbalance 
between buyer- and seller-initiated trades that took place 
in the specified time interval. 
For that matter, the order flow 
is also known as order imbalance 
in the market microstructure literature 
(see, for example, \cite{CF:2000}).
Order flow is an important measure of net trading pressure for a tradable asset.

For the purpose of modeling, 
we postulate that the order flow $X_t$ 
is a mean-reverting process.
Denoting the standard Brownian motion by $w_t$,
the stochastic process of $X_t$ is
\be 
\label{e:X_t}
\dr X_t = c(m - X_t)\dr t + \eta\, \dr w_t.
\ee
Here, $m$ is the long-run average, $c$ is the speed of mean reversion, 
and $\eta$ is the volatility of order flow $X_t$.
For any given day or trading session, 
these parameters are assumed to be constants.
Evidence of mean reversion and the estimates of these parameters from market data
will be provided subsequently in Section \ref{Empirical Study}.

The trade price $P(X_t, t)$ at the end of the 1-minute interval, time $t$,
differs from a corresponding unobservable price $S(t)$.
In \cite{Roll:1984}, $S(t)$ is called the true price, 
which captures the ``value'' of an asset.
The model of \cite{HS:1997} calls $S(t)$ the unobservable fundamental value.
In \cite{MRR:1997}, $S(t)$ is the post-trade expected value
of the security and its relationship with the trade price $P(X_t, t)$ is---at 
the transaction level with parameter $\theta$---as follows:
\be
P(X_t, t) = S(t) + \theta\, \hbox{sign}(X_t) + \hbox{i.i.d. noise}.
\ee
More recently, \cite{FT:2015} extend it to examine
the informativeness of trades and quotes in the FTSE 100 index futures market.

\cite{CJP:2004} provide the multiplicative form of this relationship
with a generic and unspecified price impact function $f(X_t)$. 
Namely,
\be
P(X_t, t) = S(t) \exp\big( f(X_t) \big).
\label{e:P(X_t, t)}
\ee
They call $P(X_t, t)$ the supply curve and $S(t)$ 
is assumed to be a semi-martingale.
In particular, $f(X_t)$ is taken to be a function with the following properties:
non-decreasing in $X_t$; twice differentiable; and  $f(0) = 0$.
Given these properties, it follows that
$f(X_t) > 0$ for $X_t > 0$, and $f(X_t) < 0$ for $X_t < 0$.

The main objective of \cite{CJP:2004} is to extend 
the first and second fundamental theorems of asset pricing.
They formulate a new model that takes into account
the fact that trading liquidity is not infinite,
which is implicitly assumed in the classical asset pricing models.
As an example of their fundamental theory, 
they consider an extension 
of the Black-Scholes economy that incorporates liquidity risk.
The functional form of the price impact function $f(x)$
in the extended Black-Scholes economy is \emph{assumed} to be linear,
i.e., $f(x) = \alpha x $, where $\alpha$ is a positive constant. 

Starting from \eqref{e:P(X_t, t)},
our main objective is to derive the functional form of $f(x)$ 
by introducing the market price of liquidity,
within the framework of \cite{CJP:2004}.
Our model allows us to derive not only the price impact function
but also the implications for the stochastic nature of volatility.
We also obtain a measure of market depth that depends on
all the parameters of the market price of liquidity risk.
 
For notational convenience, henceforth we write $P(X_t, t)$ 
as $P_t(X)$ or simply $P_t$, and $S(t)$ as $S_t$.
It is imperative to highlight that the order flow in our model is not tick-by-tick
but aggregated over the regular time interval of 1 minute chronicled by time $t$.
This is different from the analysis framework of structural models
such as \cite{MRR:1997}, which examines how ``surprises'' in the tick-by-tick
signed trades affect the true price---which occur at irregular time interval---and 
the structural parameters are estimated
with a linear specification of price change.

In addition to \eqref{e:X_t}, we assume that the unobservable true price $S_t$
is a geometric Brownian motion.
We denote the drift rate of $S_t$ by $\mu_S$, 
the volatility by $\sigma_S$, and with $z_t$ being the standard Brownian motion,
the model for $S_t$ is
\be
\frac{\dr S_t}{S_t} = \mu_S\, \dr t + \sigma_S\,\dr z_t.
\label{e: dSt over St}
\ee

In general, given two standard Brownian motions $w_t$ and $z_t$,
there is no a priori reason to assume that they are uncorrelated.
Therefore, with correlation coefficient $\rho$, 
which is taken to be a constant, we write
\be
\dr w_t\, \dr z_t= \rho\, \dr t .
\ee
Due to this correlation, it follows that the covariance 
between the true price return and change of order flow is
\be
\frac{\dr S_t}{S_t} \dr X_t = \rho \eta \sigma_S \dr t .
\label{e: SX correlation}
\ee
It is important to highlight that $\mu_S$, $\sigma_S$, and $\rho$
cannot be estimated, as the true price process \eqref{e: dSt over St} is unobservable.

Even though the true price $S_t :=S(t)$ does not depend on the order flow directly,
it is related to the order flow volatility $\eta$ in our framework.
Suppose $\rho=1$, i.e., perfect correlation. 
If $\dr X_t > 0$, i.e., more buying pressure, \eqref{e: SX correlation} suggests that
the true price $S_t$ will experience an additional upward drift
of an amount $\eta \sigma_S$ per unit of order flow on average.
Conversely, if  $\dr X_t < 0$,
the true price tends to drift downward by $\eta \sigma_S$ for each unit of order flow.
In general, so long as $\rho \ne 0$, the true price 
is indirectly affected by the order flow.
Being mindful that correlation is not causation,
\eqref{e: SX correlation} nevertheless can be interpreted as
some sort of indirect ``permanent'' price impact of order flow on the true price.
The degree by which order flow volatility $\eta$ 
and the unobservable volatility $\sigma_S$ impacts the true price
is moderated by $\rho$.
Only in the  special case of $\rho = 0$
does the model exclusively deal with the temporary price impact.

Therefore, our modeling framework contains 
elements of temporary and permanent price impact.
The former has no impact on the true price.
The latter impacts $S_t$ indirectly through the order flow volatility $\eta$ 
and the unobservable volatility  $\sigma_S$.
In a sense, $\rho\eta\sigma_S$ can be interpreted as the extent or degree by which 
information in the order flow affects the true price.
This interpretation is consistent with the finding of \cite{CHST:2019}.
They propose that the volatility of order flow is 
a proxy for the cost of adverse selection, 
and empirically they find that order flow volatility is significantly higher 
prior to important announcements that are likely to elevate information asymmetry.

We call $f(X_t)$ the price impact function of order flow $X_t$.
The trade price $P_t$ deviates from the true price $S_t$
due to the price impact of order flow.
When an investor submits a market order to trade,
the price at which the order is executed is $P_t$. 
Being a function of the order flow $X_t$,
the factor $\xi(X_t) := \exp\big( f(X_t) \big) > 0$ 
accounts for the price difference
between $P_t$ and the unobservable $S_t$.
With respect to the true price $S_t$, 
investors who demand liquidity have to brace themselves
for the price impact risk for which $\xi(X_t) \ge 1$
if $X_t$ has more buying order flow than selling order flow, 
and $\xi(X_t) \le 1$ if the reverse scenario occurs.
The notion that $P_t \ne S_t$ is a form of market friction.

When $X_t = 0$ during a trading session, 
it could be that the buy and sell orders
are perfectly balanced at time $t$.
We have the ``boundary condition'' as follows:
\be
\xi(0)=1 \quad \Longleftrightarrow \quad f(0)=0.
\ee
It is rare for the order flow of buys and sells to be exactly balanced, 
so that $X_t = 0$ during the trading session.
But after the trading hours, no trading occurs and $X_u = 0$ for $u > T$,
where $T$ is the closing time.
It is important to highlight that this case is fundamentally different.
Obviously, when the trading session is over,
$P_t$ is the last traded price ($t \le T$), 
and it remains constant until the next trading session.
The relationship between the constant $P_t$ and the unobservable $S_u$
cannot be modeled by \eqref{e:P(X_t, t)} anymore.
In other words,  the framework of \cite{CJP:2004}, \eqref{e:P(X_t, t)}, implicitly
assumes that the asset is being transacted during the trading hours.

\section{Analysis with It\^{o}'s Calculus}  \label{Gen Analysis}

From \eqref{e:P(X_t, t)}, we have
\be
\dr \log P_t = \dr \log S_t + \dr f(X_t).
\label{e:d logP_t}
\ee
By applying It\'{o}'s formula, 
and in view of the mean-reverting dynamics of the order flow
\eqref{e:X_t}, we obtain
\bea \label{eq2.1}
\dr \log P_t \is
\left[ \mu_S -\frac{1}{2}\sigma^2_S 
+ c(m-X_t)g(X_t) + \frac{1}{2} \eta^2 g^{\prime}(X_t) \right] \dr t \nonumber  \\
& & \hspace{2em} + \sigma_S\, \dr z_t + \eta g(X_t)\, \dr w_t,
\eea
where we have defined the gradient of the price impact function as follows:
\be
g(X_t) := f^\prime(X_t).
\ee
Note that $g(X_t) \ge 0$ since $f(X_t)$ 
is assumed to be non-decreasing in $X_t$.
Also, \eqref{eq2.1} suggests that the gradient $g(x)$ of the price impact function,
when modulated with the mean reversion of order flow,
provides an extra impetus for the logarithmic price to change.
Note also that the variance of the order flow, $\eta^2$,
contributes positively to the drift rate as well.

In view of the stochastic terms in \eqref{eq2.1}, i.e., 
$\sigma_S\, \dr z_t + \eta g(X_t)\, \dr w_t$,
it is natural to consider the total variance denoted by $ \sigma_{P}^2(x)$.
It comprises the variance $\sigma_S^2$ from the standard Brownian motion $\dr z_t$,
the variance $\eta^2 g(X)^2$ arising from the order flow randomness,
and the covariance $\rho \eta \sigma_S g(X_t)$ between these two stochastic processes.
That is, the variance of the logarithmic return \eqref{eq2.1} is obtained as
\be \label{eq2.2}
\sigma_P^2(X_t) := \sigma_S^2 + \eta^2 g^2(X_t) + 2\rho \eta \sigma_S g(X_t) .
\ee
It follows that the logarithmic return \eqref{eq2.1}, by It\^{o}'s formula,
is connected to the following stochastic differential equation for the simple return
based on the trade price $P_t$:
\be \label{eq2.3}
\frac{\dr P_t}{P_t} = \mu_P(X_t)\, \dr t + \sigma_S\, \dr z_t + \eta g(X_t)\, \dr w_t,
\ee
where 
\be
\mu_P(X_t)= \mu_S + \big\{ c(m-X_t) + \rho  \eta \sigma_S  \big\} g(X_t) 
+ \frac{1}{2} \eta^2 \big\{ g^{\prime}(X_t) + g^2(X_t) \big\} .
\label{e:mu_p(x)}
\ee
The process \eqref{eq2.3} is under the physical measure $\pp$.
So are the order flow process \eqref{e:X_t} 
and the true price process \eqref{e: dSt over St}.

A few remarks are in order.
First, the simple return \eqref{eq2.3} 
is the result of stochastic calculus applied to the assumptions
in Section~\ref{sect:Model Setup and Motivation}.
It has two sources of risk.
One is due to the Brownian fluctuation of the true price process,
whose volatility $\sigma_S$, though unobservable, is assumed to be constant.
The other source is from the order flow process.
The volatility $\eta g(X_t)$ in \eqref{eq2.3} is dependent on $X_t$,
which means that the volatility $\sigma_P (X_t)$ of the simple return is stochastic,
as in \eqref{eq2.2}.

Next, if the correlation coefficient $\rho$ is non-negative,
i.e., $\rho \ge 0$, then $\sigma_{P}(X_t) > \sigma_S$,
implying that trading itself creates an additional variance
that is quadratically related to the volatility $\eta$ of the order flow.
Everything else being equal,
a larger $\eta$ will bring about a larger volatility in the simple return,
which is consistency with intra-day market reality.
On the other hand,
when the trading session is over, there is no order flow, 
and $P_t(0)$ is the last traded price,
which does not change anymore until the next trading session.
Thus, $\sigma_P(0) = 0$ after the trading session.

Finally, it is clear from \eqref{e:mu_p(x)} that
the drift rate is also stochastic,  mean reverting in $X_t$, 
and nonlinearly dependent on order flow volatility,
which is consistent with the empirical findings of \cite{EL:2002}.
In all of these, the gradient of the impact function of order flow is the main driver
that makes the volatility $\sigma_P(X_t)$ 
and the drift rate $\mu_P(X_t)$ of the physical price process stochastic,
since $X_t$ is a stochastic process according to \eqref{e:X_t}.

\subsection{Market prices of risks}
Now, let $\lambda_t^z := \lambda^z(X_t)$ 
and $\lambda_t^w := \lambda^w(X_t)$ 
denote the market prices of risks associated with $z_t$ and $w_t$, respectively.
In general, $\lambda_t^z$ and  $\lambda_t^w$ can depend on the true price $S_t$.
But for a start, we assume their dependence on the order flow only.
To carry out a change of probability measure, we define
\be 
\dr \wtil z_t := \dr z_t + \lambda_t^z \dr t \qquad \hbox{and} 
\quad
\qquad \dr \wtil w_t := \dr w_t + \lambda_t^w \dr t .
\label{e:G-transform} 
\ee

As we shall see later,
$\lambda_t^z$ is the market price of market risk in classical asset pricing.
The novel construct $\lambda_t^w$, on the other hand,
has its origin in the liquidity uncertainty arising from order flow fluctuation.
Henceforth, we call $\lambda_t^w$ the market price of liquidity risk.

Let $\Q$ be a probability measure such that the stochastic processes 
$\wtil z_t$ and $\wtil w_t$ 
are standard Brownian motions.
Substituting \eqref{e:G-transform} into (\ref{eq2.3}), 
it is straight-forward to obtain
\be \label{eq2.5}
\frac{\dr P_t}{P_t} = \big( \mu_P(X_t) - \sigma_S \lambda^z(X_t) 
- \eta g(X_t) \lambda^w(X_t) \big) \dr t 
+ \sigma_S\, \dr \wtil z_t + \eta g(X_t) \dr \wtil w_t
\ee
under the risk-neutral measure $\Q$.

Suppose that price manipulation is not possible for $P_t$. 
In this case, we can assume that there is no arbitrage opportunity in the market, 
which is equivalent to saying that, under the risk-neutral measure $\Q$, 
the denominated trade price $P_t/B_t$ is a martingale, 
where $B_t=\e^{rt}$, and $r$ denotes the risk-free spot interest rate. 
It follows from (\ref{eq2.5}) that a condition has to be satisfied.
Namely, 
\be \label{eq2.6}
r = \mu_P(x) - \sigma_S \lambda^z(x) - \eta g(x) \lambda^w(x) .
\ee

Next, following the standard method of It\^{o}'s calculus,  
the market price of market risk is given by
\be \label{eq2.7}
\lambda^z (x) = \frac{\mu_S - r +\kappa(x)}{\sigma_S}, 
\qquad \kappa(x) \geq 0, 
\ee
for some non-negative function $\kappa(x)$. 
Observe that if $\kappa(x)=0$,
then $\lambda^z(x)$ is independent of the order flow $x$
and the ratio $\ds \frac{\mu_S-r}{\sigma_S}$ corresponds
to the Sharpe ratio of classical asset pricing models.

In general, the function $\kappa(x)$ may be interpreted
as an order-flow-dependent gain that informed traders 
are getting when they trade strategically in a way
that maximizes their profits in the setting of \cite{Kyle:1985}.
This interpretation of $\kappa(x)$ becomes apparent
in the linear solution that we shall derive in the next section.

Combining \eqref{e:mu_p(x)}, (\ref{eq2.6}), and (\ref{eq2.7}), 
we obtain an ordinary differential equation (ODE) satisfied by $g(x)$, i.e., 
\be \label{eq2.8}
0 = - \kappa(x) + \{ c(m-x) + \rho \eta \sigma_S - \eta \lambda^w(x) \} g(x) 
+ \frac{1}{2} \eta^2 \{ g^{\prime}(x) + g^2(x) \} .
\ee
Note that the ODE (\ref{eq2.8}) involves the market price of liquidity risk 
$\lambda^w(x)$ associated to the order flow $X_t$. 

\subsection{Inhomogeneous Bernoulli Differentiation Equation}
To render \eqref{eq2.8} into the canonical form,
we define 
\be \label{eq:definition of sx}
\ds s(x) := 2\frac{\kappa(x)}{\eta^2} \geq 0 ,
\ee
and a function
\be \label{eq2.9}
p(x) := \frac{2}{\eta^2} \big\{ c(m-x) + \rho \eta \sigma_S -\eta \lambda^w(x) \big\} .
\ee
Essentially, $p(x)$ is a function of order flow $x$ due to its mean reversion
and the market price of liquidity risk $\lambda^w(x)$. 
It follows from (\ref{eq2.8}) and (\ref{eq2.9}) that 
\be \label{eq2.10}
0 = - s(x) + g^\prime (x) + p(x) g(x) + g^2(x),
\ee
which is an inhomogeneous Bernoulli differential equation\footnote{
The ODE with $s(x)=0$, i.e.,
\[ 
0 = g^\prime (x) + p(x) g(x) - q(x) g^n(x) ,\qquad n\ne 1, 
\]
is called the Bernoulli equation whose solution is known:
\[ 
g(x) = \left[ \dfrac{(1-n)A \ds\int \e^{(1-n) \ds\int p(x) \dr x+ B }q(x)\dr x + B}
{A \e^{(1-n) \ds\int p(x) \dr x} \dr x } \right]^{\dfrac{1}{1-n}}, 
\] 
where $A$ and $B$ denote the constants of integration.}.
The solution of (\ref{eq2.10}) with $s(x) \not\equiv 0$ is in general unknown.
In principle, however, (\ref{eq2.10}) can be solved numerically in conjunction
with some appropriate boundary conditions. 

As a summary, given the solution $g(x)$ of (\ref{eq2.10}), 
we have the following price impact models.
Namely, under the risk-neutral measure $\Q$, we have
\be \label{eq2.11}
\left\{ \begin{array}{l} \ds
\frac{\dr P_t}{P_t} = r\, \dr t + \sigma_S\, \dr \wtil z_t 
+ \eta g(X_t) \dr \wtil w_t , \vspace{0.1cm} \\
\dr X_t = \{ c(m - X_t) - \eta \lambda^w(X_t) \} \dr t + \eta\, \dr \wtil w_t , \\
\dr \wtil z_t \dr \wtil w_t =\rho\, \dr t .
\end{array} \right. 
\ee

\clearpage
\noindent
Moreover, in view of \eqref{eq2.8}, we rewrite $\mu_P(X_t)$ to obtain
\be \label{eq2.12}
\left\{ \begin{array}{l} \ds
\frac{\dr P_t}{P_t} = \big(\mu_S + \kappa(X_t) + \eta g(X_t) \lambda^w(X_t) \big) \dr t 
+ \sigma_S\, \dr z_t + \eta g(X_t)\, \dr w_t , \vspace{0.1cm} \\
\dr X_t = c(m - X_t) \dr t + \eta\, \dr w_t , \\
\dr z_t \dr w_t =\rho\, \dr t , 
\end{array} \right. 
\ee
under the physical measure $\pp$. 
Again we note that the volatility is stochastic through $\eta g(X_t)$,
which is dependent on the stochastic behaviors of the order flow $X_t$.

Moreover, from (\ref{eq2.12}), under the physical measure $\pp$,
the correlation between the stochastic differentials can be expressed as
\be \label{eq.cor}
\frac{\dr P_t}{P_t} \dr X_t =  \big(\rho \eta \sigma_S  + \eta^2 g(X_t) \big) \dr t.
\ee
In comparison to \eqref{e: SX correlation}, 
we have an additional term $\eta^2 g(X_t)$, 
which is attributed solely to the order flow.
Thus, our model suggests that the correlation between the simple return $\dr P_t/P_t$
and the change in order flow $\dr X_t$ is also stochastic,
since $g(X_t)$ is a function
of the order flow process, which is stochastic.

\section{Analysis Leading to Price Impact Functions} 
\label{Analysis Leading to Price Impact Functions}
To solve the ODE (\ref{eq2.10}), 
suppose $\alpha(x)$ is a special solution of (\ref{eq2.10}), i.e.,
\be \label{eq3.1}
\alpha^\prime (x) + p(x) \alpha(x) + \alpha^2(x) = s(x).
\ee 
We define
\be \label{eq3.2}
\phi(x) := \e^{-\ds\int_0^x \beta (y) \dr y} \quad \hbox{and} 
\quad \Phi(x) := \int_0^x \phi(y) \dr y , 
\ee
for some $\beta(x)$, where the usual integration rule 
$\ds\int_0^x h(y)\dr y=- \ds\int_x^0 h(y)\dr y$ applies. 

\begin{prop} \label{prop3.1}
Given that $\alpha(x)$ is a special solution,
a general solution of the ODE (\ref{eq2.10}) is obtained as follows:
\be 
g(x) = \alpha(x) + \frac{A \phi(x)}{A \Phi(x) + B} 
\quad\hbox{and}\quad \beta(x) = p(x) + 2\alpha(x), 
\label{Equation: Proposition 1}
\ee
where $A$ and $B$ are integration constants. 
\end{prop}

Proof:\:
Note from (\ref{eq3.2}) that $\phi ' (x) = - \beta(x) \phi(x)$. 
It follows that
\be
g' (x) = \alpha ' (x) - \frac{A \beta (x) \phi(x) \Gamma(x) + A^2 \phi^2(x)}{\Gamma^2(x)},
\quad \hbox{where}\quad 
\Gamma(x) :=A \Phi(x) + B . 
\ee
Also, we have
\be
p(x) g(x) = p(x) \alpha(x) + \frac{A p(x) \phi(x) \Gamma(x)}{\Gamma^2(x)} 
\ee
and
\be
g^2(x) = \alpha^2(x) + \frac{2A \alpha (x) \phi(x) \Gamma(x)}{\Gamma^2(x)} 
+ \frac{A^2 \phi^2(x)}{\Gamma^2(x)} . 
\ee
Summing these terms, we obtain 
\be
\alpha^\prime (x) + p(x) \alpha(x) + \alpha^2(x) 
+ \frac{A \phi(x) \{ -\beta(x) + p(x) + 2\alpha(x) \} }{\Gamma(x)} , 
\ee
which is equal to $s(x)$, when $\alpha(x)$ is a special solution (\ref{eq3.1}) 
and $\beta(x) = p(x) + 2\alpha(x)$, 
thus proving the proposition. 
\qed

\bigskip
Now, a boundary condition is given by $f(0)=0$, i.e., no-trade-no-impact condition.
The other boundary condition is that the slope of $f(x)$ at $x=0$, i.e.,
$g(0)=\ell$ is a positive value.
Denote $\alpha := \alpha(0)$.
From \eqref{Equation: Proposition 1} in proposition \ref{prop3.1}, we have
\be 
\ell = g(0) = \alpha + \frac{A}{B}.
\ee
Also, since $g(x) = f^\prime(x)$, by integration, 
we obtain the price impact function:
\be
f(x) = \int_0^x \alpha(y) \dr y + \log \big[ A\Phi(x) + B \big],
\ee
which implies $0 = \log B$, i.e., $B=1$.
Thus, the general price impact function is given by
\be \label{eq.sol}
f(x) = \int_0^x \alpha(y) \dr y + \log \big[ (\ell - \alpha) \Phi(x) + 1\big], 
\ee
where $\Phi(x)$ is defined in (\ref{eq3.2}). 

To obtain an explicit form of the price impact function,
we need to know the explicit form of the special solution $\alpha(x)$.
Moreover, we also need to specify the market price of liquidity risk,
$\lambda^w(x)$, which appears in the $p(x)$ function \eqref{eq2.9}.
As an analogy to the Sharpe ratio, we write the market price of liquidity risk as
\be
\lambda^w(x) = \frac{\delta-\tau x}{\eta} ,
\ee
where $\delta$ and $\tau$ are constants.
This specification suggests that 
the market price of liquidity risk $\lambda^w(x)$ is a linear function of $x$,
namely, 
\be
\eta \lambda^w(x) = - \tau x + \delta.
\label{Equation: eta lambda}
\ee
In the following, we consider two special cases to obtain solutions in closed form.

\subsection{$p(x)=p$ and $s(x) = s$}
Suppose $\tau$ in \eqref{Equation: eta lambda} 
is set equal to the speed of mean reversion $c$.
In this special case where $\tau=c$, we see from \eqref{eq2.9}
that $p(x)$ becomes constant, i.e. $p(x) = p$,
where 
\be
\ds p := \frac{2}{\eta^2}\big( {cm + \rho\eta\sigma_S + \delta}\big) .
\ee
To solve the ODE  (\ref{eq3.1}),
we conjecture that the special solution $\alpha(x)$ in this ODE is also a constant,
i.e.,  $\alpha(x)=\alpha >0$.
Consequently, the ODE (\ref{eq3.1}) becomes a quadratic equation
\be
p \alpha + \alpha^2 =s .
\ee
The solution is given by
\be 
\alpha = \frac{-p +\sqrt{p^2 + 4s}}{2} , \qquad s > 0 . 
\ee 
Moreover, from \eqref{Equation: Proposition 1} and by the setting of $p(x) = p$,  
$\beta(x)$ is also a constant $\beta$.
In this special case, $\beta$ is given by
\be 
\beta = p + 2\alpha =\sqrt{p^2 + 4s} >0 . 
\ee 

Consequently, from (\ref{eq3.2}), we have 
\be
\phi(x) = \e^{- \beta x} \quad\hbox{and}\quad \Phi(x) 
= \frac{1}{\beta}\big(1- \e^{- \beta x} \big). 
\ee
It follows from (\ref{eq.sol}) that the price impact function 
in this special case is obtained as
\be \label{eq3.5}
f(x) = \alpha x 
+ \log \left[1 + \frac{\ell -\alpha}{\beta}\big(1 -  \e^{- \beta x}\big) \right],
\ee
which is a sum of linear and  nonlinear price impact functions. 

Since $g(x) = f'(x)$, we obtain
\be
g(x) = \alpha + \frac{A \beta \e^{- \beta x}}{ 1+ A - A\e^{- \beta x}} , 
\ee
where, in light of \eqref{eq3.5}, we define $\ds A := \frac{\ell-\alpha}{\beta}$ 
to ease the notations.
Moreover,
\be
g ' (x) = f^{\prime\prime} (x)
= - \beta^2 A(1+A) \frac{\e^{- \beta x} }{[1+ A - A\e^{- \beta x}]^2} .
\ee
Hence, the nonlinear component of $f(x)$ 
is strictly concave as long as $A>0$, i.e., $\ell > \alpha$.

However, for the logarithm in \eqref{eq3.5} to be well defined,
the applicability of this price impact function
is restricted to
\be
x>-\frac{1}{\beta} \log \left( 1+\frac{\beta}{\ell -\alpha}\right).
\ee
Therefore, for \eqref{eq3.5} to be applicable to all $x$,
we have to set $\ell = \alpha$.
Consequently, the nonlinear term in \eqref{eq3.5} vanishes
and the price impact function reduces to 
\be
f(x)=\alpha x .
\label{Equation: Linear Price Impact Function}
\ee
This result suggests that we have derived a linear price impact function
that is consistent with the theoretical finding of linearity by \cite{Kyle:1985}.
In other words, our $\alpha = \ell$ is Kyle's illiquidity measure $\lambda$.

Moreover, since $\ds s = 2 \frac{\kappa}{\eta^2}$ and $p\alpha + \alpha^2 = s$,
we can write the expected return $\kappa$ of the informed trader as
\be
\kappa = \frac{\eta^2}{2} \big(p \alpha + \alpha^2\big) .
\ee
Since $\alpha < 1$ is a small number, we can approximate $\kappa$ by the leading order,
i.e., $\ds \kappa \approx \frac{\eta^2}{2} p \alpha $, 
and it follows that $\kappa$ is proportional to $\eta^2 \alpha$.
Now, in \cite{Kyle:1985}, $\lambda$, which corresponds to our $\alpha$,
is inversely proportional to the volatility of the order flow $\eta$.
Therefore, $\kappa$ is directly proportional to $\eta$.
Interestingly, the expected profit or return of the informed trader in \cite{Kyle:1985}
is also directly proportional to $\eta$,
which provides a justification for naming $\kappa$ as the expected return
of the informed trader in the earlier section.

\subsection{$p(x) = p + qx$ and $s(x)=0$}
Now, we consider the case where all traders are either more or less equally informed,
or more or less equally uninformed.
This case can be modeled by setting $s(x) = 0$, which is as good as
setting $\kappa(x) = 0$ according to \eqref{eq:definition of sx}.

As alluded to in Section \ref{Introduction},
some empirical findings such as \cite{HT:2008} suggest that the impact function $f(x)$ 
is convex until some $x <x^*$ and concave afterwards.
It turns out that this S-shape feature can be modeled 
in our framework by setting $p(x) = p + qx$ and $s(x) = 0$.
To obtain an analytical solution, we conjecture that the special solution
$\alpha(x)=0$, so that, according to \eqref{Equation: Proposition 1},
\be
\beta(x)=p(x) .
\ee
Also, note from the definition \eqref{eq:definition of sx}
that in this case, $\kappa(x)$ in \eqref{eq2.7} is zero.
Therefore, this special case corresponds to modeling the
effect of illiquidity in the classical framework of asset pricing
characterized by the Sharpe ratio.
 
From Proposition \ref{prop3.1}, 
a general solution to the ODE (\ref{eq2.10}) with $s(x)=0$ is obtained as
\be 
g(x) = \frac{A \phi(x)}{A \Phi(x) + B}, \qquad \phi(x) = \e^{-\ds\int_0^x p(y) \dr y} ,
\quad \Phi(x) = \int_0^x \phi(y) \dr y . 
\label{Equation: g(x)}
\ee
Accordingly, from (\ref{eq.sol}), we obtain a closed form solution:
\be \label{eq3.6}
f(x) = \log \big[ 1+ \ell \Phi(x) \big],
\qquad g(x) = \frac{\ell \phi(x)}{1+ \ell \Phi(x) }, \qquad x\in I, 
\ee
where $I$ denotes the interval on which the impact function $f(x)$ is well defined, 
i.e., $1+ \ell \Phi(x) >0$ for $x\in I$. 

\bigskip
\begin{prop} \label{s(x)=0}
Suppose that $p(x)$ is increasing in $x$ 
and it has a unique solution $x^* \in I$ for $p(x)=0$. 
Then, the impact function $f(x)$ is convex for $x <x^*$ 
and concave for $x > x^*$.
\end{prop}

Proof:
From (\ref{eq3.2}), since $\phi '(x) = - p(x) \phi(x)$, it is readily seen that 
$g'(x) = f^{\prime\prime}(x)$ is given by
\be
g ' (x) =  - \ell \phi(x) \frac{ p(x) [1+ \ell \Phi(x) ] 
+ \ell \phi (x) }{[1+ \ell \Phi(x) ]^2}.
\ee
We define
\be
h(x) := p(x) [1+ \ell \Phi(x) ] + \ell \phi (x),
\ee
and we have
\be
h '(x) = p ' (x) \big[1 + \ell \Phi(x) \big].  
\ee
Since $1+ \ell \Phi(x) >0$ for $x\in I$, $h(x)$ is increasing, $h(x)<0$ 
for $x <x^*$ and $h(x)>0$ for $x > x^*$. 
Moreover, since $\ell \phi(x) >0$, we have $g '(x) 
\geq 0$ for $x <x^*$ and $g '(x) \leq 0$ for $x > x^*$.
\qed

\bigskip

In particular, when  $p(x)= p + q x$ with $q>0$, 
we have from \eqref{Equation: g(x)} the following explicit form:
\be
\phi(x) = \e^{-p x - \tfrac{q}{2} x^2} ,\qquad 
\Phi(x) = \int_0^x \e^{-p y - \tfrac{q}{2} y^2} \dr y .
\ee
It remains to define the interval $I$ for which $1+ \ell \Phi(x) >0$ 
for this special case of linear $p(x)$.
To this end, since 
\be
p y + \frac{q}{2} y^2 = \frac{q}{2} \left( y^2 + \frac{2p}{q} y \right) 
= \frac{q}{2} \left( y + \frac{p}{q} \right)^2 - \frac{p^2}{2q} , 
\ee
we have $\ds \phi(x) = \e^{-px - \tfrac{q}{2}x^2}$ and it follows that
\be
\Phi(x) =
\int_0^x \e^{-p y - \tfrac{q}{2} y^2} \dr y 
= \e^{p^2/2q} \int_{p/q}^{x+p/q} \e^{-\tfrac{q}{2} y^2} \dr y 
= \frac{\e^{p^2/2q}}{\sqrt{q}} 
\int^{\sqrt{q}\left(x+p/q\right)}_{p/\sqrt{q}} \e^{-y^2/2} \dr y . 
\ee

Denoting the cumulative distribution function 
of standard normal distributions by $N(x)$, i.e., 
\be
N(x) := \int_{-\infty}^x \frac{1}{\sqrt{2\pi}} \e^{-y^2/2}\dr y,
\ee
we obtain
\be \label{eq3.8}
\Phi(x) = \frac{\sqrt{2\pi}\e^{p^2/2q}}{\sqrt{q}} 
\Big( N\big( \sqrt{q}(x+p/q) \big) - N\big(p/\sqrt{q}\big) \Big). 
\ee
This is essentially the Gauss error function,
which has the feature of S shape, with an inflection point at
\be
x^* = -\frac{p}{q} .
\label{Equation: Inflection Point}
\ee

In view of the result that $f(x) = \log\big[1+\ell \Phi(x) \big]$,
it can be easily seen that the impact function $f(x)$ is well defined for all $x\ge 0$,
since $\Phi(x) \ge 0$ when the order flow is not negative.
On the other hand, 
a constraint will emerge when we want $f(x)$ to be well defined for $x \to -\infty$ as well.
Since $N(-\infty) = 0$,
the constraint on the parameters ($\ell$, $p$, and $q$) is
\be \label{eq3.10}
1- \ell \sqrt{\frac{2\pi}{q}} \e^{p^2/2q} N\big(p/\sqrt{q}\big) >0.
\ee
Since $N(x)$ is monotonically increasing, 
we conclude that if  \eqref{eq3.10} is satisfied,
then the impact function $f(x)$ is well defined for all $x$,
i.e., $I = (-\infty, \infty)$, which is the set of real numbers.

In light of Proposition \ref{s(x)=0},
and knowing that the inflection point \eqref{Equation: Inflection Point}
depends only on $p$ and $q$, 
we may interpret $p(x) = p + qx$
as a function that governs how the gradient and curvature of the S-shape
price impact function vary with the order flow.
From \eqref{Equation: Inflection Point}, we see that
if $p = 0$, the price impact function will become symmetric with respect to
the origin, i.e., $x^* = 0$, implying that buyer- 
and seller-initiated trades produce symmetrical price impacts.
By using only the volume $V$ but not the trade sign,
the industry models mentioned in Section \ref{Introduction}
treat the price impacts as symmetric based on their in-house research
with aggregated data over many years.
In general, $p$ is not zero in our model,
and we interpret it as a parameter that
controls the extent of asymmetry in the price impacts of buys versus sells.

As for the $q$ parameter in $p(x)= p+ qx$, since it is the coefficient of $x$, 
it can be interpreted as the scaling factor.
From \eqref{Equation: g(x)} and \eqref{eq3.6}, we find that
\be
\phi(x) = \e^{-px -\tfrac{1}{2}\left(\sqrt{q}x\right)^2} .
\ee
Thus, if $q$ is large, the gradient of the S-shape curve
approaches zero rapidly even for a small order flow $x$.
Conversely, if $q$ is small, it takes big order flows to start to reduce
the gradient of the price impact function.
We could then interpret $q$ as a scaling parameter
that regulates how the S-shape $f(x)$ spreads out with respect to $x$
and how its gradient starts to become smaller and smaller.

As a matter of fact,
the square root function in the industry models does not have a
scaling parameter\footnote{
Though the volume $V$ is scaled by EDV (expected daily volume),
it is either fixed or probably estimated by exogenous methods.
}.
This could be a reason why the square root price impact function
may not be able to model the price impact well
when it starts to taper off for larger order flows.
 
We now consider the linear market price of liquidity risk 
\eqref{Equation: eta lambda}, i.e.,
\be
\eta \lambda^w (x) = -\tau x + \delta, \quad\hbox{such that}\quad 
\tau > c >0 .
\label{e:market price of liquidity risk}
\ee
In other words, 
$\lambda^w (x)$ is negative for buyer-initiated trades
and positive for seller-initiated trades. 
There is an intuitive interpretation of this property.
Under the risk neutral measure, 
the stochastic process of order flow in \eqref{eq2.11}
contains $-\eta\lambda^w(x) = \tau x - \delta$ in the drift.  
Thus, a positive order flow will generally increase the drift rate, and vice versa.
Accordingly, $\lambda^w(x)$ could be interpreted as a risk-neutral adjustment 
to the order flow to reflect the positive feedback of order flow.

On the other hand, under the physical measure, 
the simple return \eqref{eq2.12} contains 
$\lambda^w(X_t)$ in the drift rate, i.e.,
$
\mu_S  -  g(X_t) \big( \tau X_t - \delta\big) .
$
since $\kappa(X_t) = 0$.
At a small time scale (such as 1-minute interval),
\eqref{eq2.12} can be discretized as
\be
P_{t+1} = P_t \big[ 1 + \mu_S  -  g(X_t)\big(\tau X_t-\delta\big)\big] 
+ \hbox{noise terms}.
\ee
We see that a positive order flow larger than $\ds \frac{\delta}{\tau}$
will be followed by a price reduction,
and conversely for negative order flow.
In other words, it must have been that $P_t$ is higher (lower) by about 
$g(X_t) \big(\tau X_t-\delta\big)$
for buy (sell).
Accordingly, liquidity consumers pay the market price of liquidity risk
to liquidity providers.
Put differently, market makers and liquidity providers expect to 
charge this amount for rendering their service.

Now, by the definition of (\ref{eq2.9}) 
and given the specified market price of liquidity risk 
\eqref{e:market price of liquidity risk}, we have 
\be 
\label{e:specific px}
p(x) = \frac{2}{\eta^2} [ c m + \rho \eta \sigma_S - \delta ] + \frac{2}{\eta^2} 
\big(\tau -c\big) x,
\ee
which implies that 
\be \label{p and q}
p := p(0) = \frac{2}{\eta^2} \big[ c m + \rho \eta \sigma_S -\delta \big]
\quad\hbox{and}\quad q :=\frac{2}{\eta^2} \big(\tau -c\big)>0 .
\ee
We see that the constituent parameter $p$ in \eqref{p and q} when $x=0$ 
consists of a few parts:
$cm$ comes from the mean reversion property of the order flow;
$\rho \eta \sigma_S$ is 
the covariance between true price's log return and the order flow;
and $\delta$ comes from the linear market price of liquidity risk.
Thus, we see that the volatility intrinsic to the asset return, $\sigma_S$,
the volatility of order flow arising from trading, $\eta$,
their correlation $\rho$, the mean reversion parameters, $c$ and $m$,
as well as the market price of liquidity risk $\delta$,
are important factors of price impact.
As for the $q$ coefficient in \eqref{p and q},
it is related to $\tau$ in the market price of liquidity risk.
It follows from \eqref{Equation: Inflection Point} that the inflection point
can be expressed as
\be
x^* = \frac{\delta - cm - \rho\eta\sigma_S}{\tau-c} .
\label{Equation: Inflection Point 2}
\ee
It encapsulates the ingredients of two defining parameters ($\delta$ and $\tau$)
in the market price of liquidity risk\footnote{
Only $c$, $m$, and $\eta$ can be estimated from the time series of
order flows.
The other parameters, namely, $\delta$ and $\tau$ of the market price of liquidity risk,
as well as $\rho$ and $\sigma_S$ for the true price process, however, 
are not observable.
}.
In our empirical study (Section \ref{Empirical Study}), 
we shall discuss why the inflection point
may be interpreted as the market depth.

As a summary, 
we have derived a price impact function \eqref{eq3.6} with \eqref{eq3.8}
from the notion of market price of liquidity risk.
The function has only three parameters,  $\ell$, $p$, and $q$.
For this price impact function, \eqref{e:P(X_t, t)} becomes, in view of \eqref{eq3.6}
\be
P_t = S_t\big[ 1 + \ell \Phi(X_t)\big].
\label{e:Kyle}
\ee
When the trade size $|x|$ is small, it can be easily seen that $\Phi(x) \approx x$,
and we recover the linear price impact function of \cite{Kyle:1985}.
Therefore, even in the nonlinear case, 
$\ell$ plays the role that is  analogous to Kyle's $\lambda$,
which is a measure of illiquidity.

\section{Empirical Study} \label{Empirical Study}
In Section \ref{Analysis Leading to Price Impact Functions},
we have derived the \cite{Kyle:1985} linear price impact function of order flow,
\eqref{Equation: Linear Price Impact Function},
by assuming that an informed trader is present.
We have also obtained a closed-form solution of a nonlinear price impact function 
\eqref{eq3.6} with \eqref{eq3.8}, \eqref{e:specific px}, and \eqref{p and q} 
from the notion of market price of liquidity risk when all traders 
are approximately equally informed.
The parameters of the nonlinear price impact function
are $\ell$, $p$, and $q$.
For clarity, 
we make it explicit that the S-shape price impact function $f(x)$ 
in our empirical study is
\be
f(x) = \log\left[ 1  + \ell \frac{\sqrt{2\pi}}{\sqrt{q}}\, \e^{\tfrac{p^2}{2q}}
\left(
\frac{1}{\sqrt{2\pi}} \int_{-\infty}^{\sqrt{q}\,\left(x + p/q\right)}  
\e^{-\tfrac{y^2}{2}} dy
- \frac{1}{\sqrt{2\pi}} \int_{-\infty}^{p/\!\sqrt{q}}  \e^{-\tfrac{y^2}{2}} dy
\right)
\right].
\label{Equation: S-shape Function}
\ee

In this section, we present evidence that our S-shape price impact function
is superior to the heuristic square-root function used in practice 
(see Section \ref{Introduction}),
for 4 different futures contracts written on the Nikkei 225 index.
The empirical study 
allows us to examine the properties of our price impact functions
in the light of data analysis results.

\subsection{Nikkei 225 Index Futures}
We first present two main reasons for choosing the Nikkei futures.
First, Nikkei 225 index is a well known stock market index
for the Japanese stock market (see, for example, \cite{AM:2017}).
According to the monthly reports compiled by the World Federation of Exchanges,
the Japan Exchange Group (JPX) was the third largest stock exchange
(after NYSE and Nasdaq) in the world\footnote{
At the time of writing, the monthly statistics are available from
\href{https://www.world-exchanges.org/home/index.php/statistics/monthly-reports}
{https://www.world-exchanges.org/home/index.php/statistics/monthly-reports}.
} during our sample period from November 2013 
to June 2017, about 3 years and 7 months.

Next, at least 4 actively traded futures contracts are written on it, 
each of different specification. 
These 4 Nikkei futures contracts provide an ideal testbed to study
the fragmentation of and the competition for order flow.
This study is made possible because the trading hours 
of these futures overlap substantially.
In particular, we focus on the Asian trading hours
when the underlying stocks of Nikkei 225 index are being traded.
Specifically, our data sets correspond to JPX's normal trading hours
from 9 AM to 3 PM Japan Standard Time, which is UTC+9:00. 
In addition to individual traders,
many hedge funds and proprietary trading firms 
are trading these futures based on strategies such as calendar spread,
quanto spread, and so on.
Their algorithmic trading activities most likely will contribute toward
providing and consuming the liquidity of these futures in the electronic markets
of Nikkei futures.

Finally, and most importantly, it is common knowledge in the market microstructure
that between the big NK futures and the mini NO futures,
the latter is expected to be more liquid.
This is because NK futures has twice the minimum tick size 
of all the other 3 futures specifications. 
Between the onshore JPX futures and the offshore CME's futures
during the Asian trading hours, the former are expected to be more liquid.
The empirical results of our models 
ought to be consistent qualitatively with these stylized facts.
In summary, these 4 Nikkei futures are selected so as to check the validity
of our model and the econometrics we employ.

\clearpage

\subsection{One-Minute Order Flows from Tick Data}
We downloaded the tick data of these 4 futures contracts from Bloomberg,  
from 4 November 2013 through 7 June 2017.
The last trading day of the front-month futures, which occurs every quarter,
is common for these 4 Nikkei contracts.
We take only the near-month futures for each quarterly maturity cycle,
since typically trading is 
sparsely
observed for the far-month futures.
Towards 
the last trading day, though, both the near- and far-month contracts
are active and both are taken as samples.

Given that trades and quotes (best bids and asks) are synchronized in time 
within the same file of records,
we can identify the best bid and ask prices
that are freshest before the trade occurs. 
That is, this pair of best bid and ask prices is the one that is closest
chronologically to the time at which the trade takes place.
According to the standard practice, we then compute the average of the best bid
and ask prices, which serves as the midpoint price.
A trade is said to be buyer (seller) initiated trade and given a trade sign of 
positive (negative) value of 1 if the trade price is strictly larger (smaller) 
than the midpoint price.
If the trade price is equal to the midpoint price, then we assign the value of zero
for its trade sign, so that it does not affect the 1-minute order flow.
On average, for each trading day,
the percentage of unsigned trades is about 23\%.

We then construct time series of 1-minute order flows along with the prices
starting from 9:00:00 to 9:00.59, 9:01.00 to 9:01.59
and so on, through 14:59.00 to 14:59:59 for all the signed trades.
Consequently, for each trading day, 
we have 360 sets of 1-minute observations.

The indigenous Nikkei 225 futures (large contracts)
and Nikkei 225 mini futures are electronically traded on the Japan Exchange Group (JPX).
In addition, Chicago Mercantile Exchange (CME) offers  Nikkei futures
settled in both yens and dollars.
These 4 futures contracts are identified by Bloomberg's ticker symbols,
as shown in Table~\ref{t:tickers}.
\[
\hbox{Table~\ref{t:tickers} is about here.}
\]

The two indigenous futures of JPX---``big'' Nikkei (NK) and 
mini Nikkei (NO) futures---are considered as onshore.
They have different contract sizes and minimum tick sizes.
By contrast, identified as NH for yen-denominated futures 
and as NX for dollar-denominated futures,
these two CME futures contracts are offshore. 
Regardless whether it is onshore of offshore, 
professional and retail futures traders 
have electronic market access to these 4 futures.

From Table~\ref{t:tickers}, note that the contract size 
and tick size of JPX's big Nikkei futures (NK) is the largest.
On the other hand, the contract size of JPX's NO,
being described as mini, has the smallest contract size
of \yen 100 for each futures point.
Therefore, the hedge ratio is one NK contract to 10 NO contracts.
From the contract size of CME's NH futures,
we see that the hedge ratio is one NH contract to 5 mini contracts.
Correspondingly, the hedge ratio is one NK contract to two NH contracts.
As for the dollar-denominated NX futures, 
the hedge ratio is determined by the spot currency exchange rate.
If the FX rate was a dollar to \yen 100,
then the hedge ratio would be one to one between NH and NX.
Otherwise, it would be 5 NX contracts to 4 NH contract 
if a dollar was worth \yen 80,
and 5 NX contracts to 6 NH contracts if a dollar could exchange for \yen 120.
These institutional characteristics are important to understand
the empirical results in this section.

We see from Table~\ref{t:tickers} that the JPX mini Nikkei futures NO
has the largest volume on daily average.
Note that it is about 10 times more than the JPX big Nikkei futures NK.
As expected, the offshore futures of CME's NH and NX are
not as actively traded during the Asian hours,
since the daily average volumes are, respectively, 7,331 contracts
and 3,703 contracts.

In terms of the average daily aggregate of positive order flows
and that of negative  order flows,
we find again the same pattern, namely, mini NO futures has the largest
numbers of contracts: 75,377 contracts of positive order flow
and 75,885 contracts of negative order flow.
On the other hand, CME's dollar-denominated NX futures has the smallest quantities
of 742 and 768 contracts, respectively,   
during Asian trading hours.
Notice that across these 4 futures contracts, 
the daily average positive and negative order flows are more or less equal.

The last two columns in Table~\ref{t:tickers} contain
the average 1-minute order flow and the volatility estimate $\widehat{\eta}$.
Over the span of 3 years and 7 months, positive and negative order flows
seem to be equally likely to occur, 
so that the daily average is statistically no different from 0 for each of
the 4 futures specifications.

\begin{figure}
\caption{\small One-Minute Order Flows of Four Futures on 15 June 2016.}
\centering
\bigskip
{\small Panel A: JPX's Nikkei 225 Large Contract (NK)}
\vskip-0.1em
\includegraphics[height=0.44\textheight]{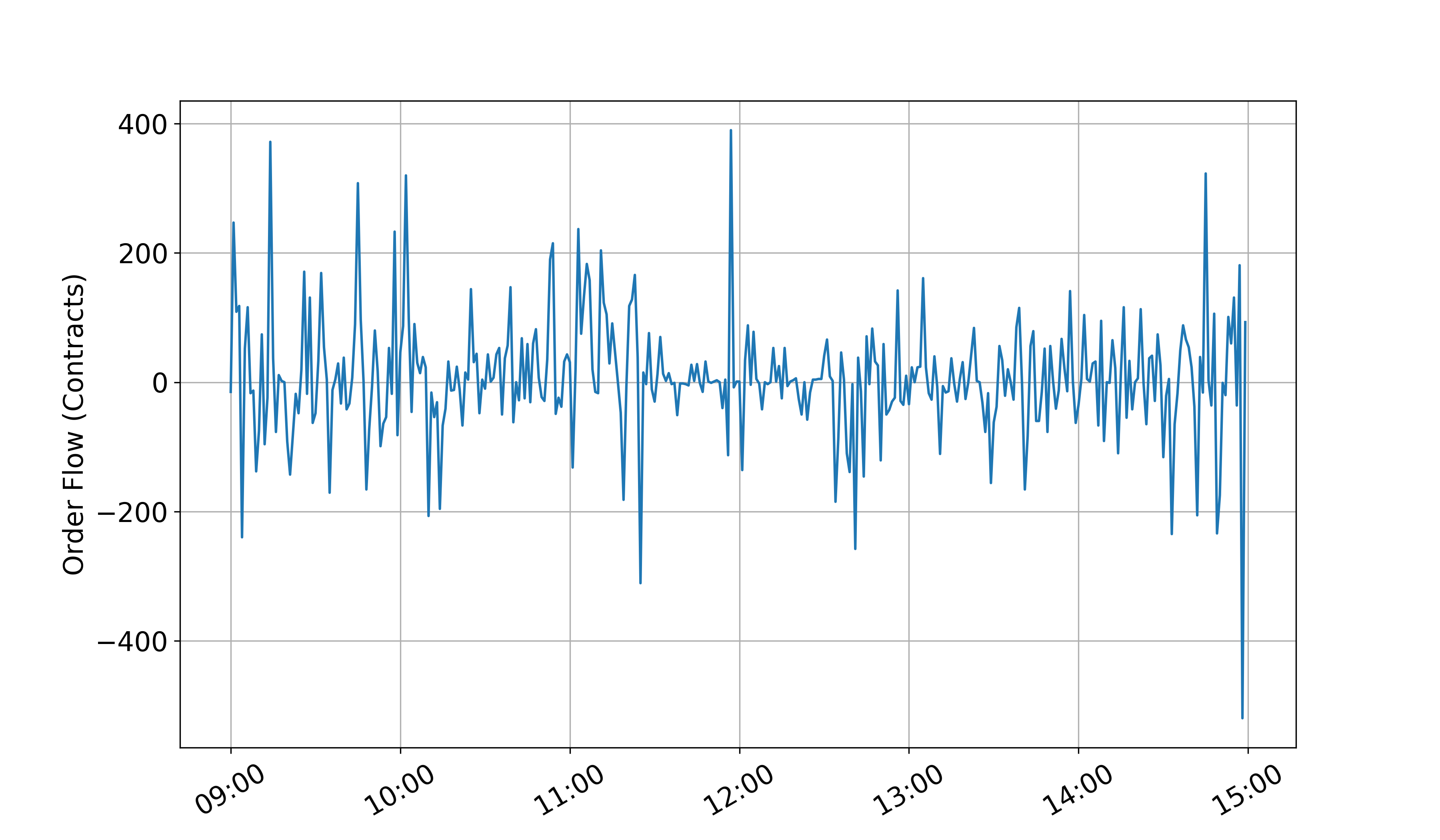}

\vspace{1em}
{\small Panel B: JPX's Nikkei 225 Mini Contract (NO)}
\vskip-0.15em
\includegraphics[height=0.44\textheight]{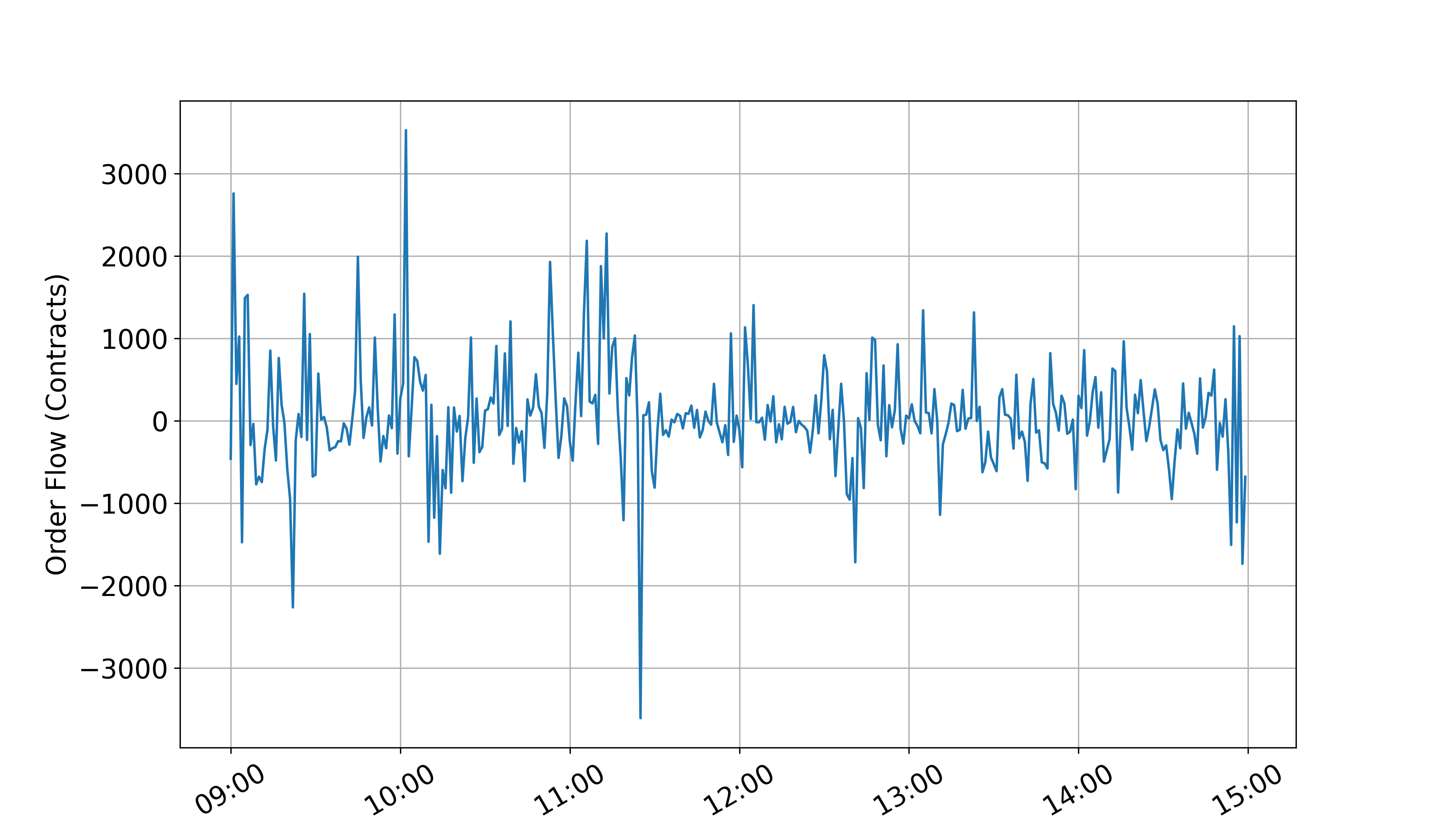}
\label{fig:Order Flows}
\end{figure}

In Figure \ref{fig:Order Flows}, we plot the intra-day time series of 1-minute order flows
from 9 AM to 3 PM Japan Standard Time on 15 June 2016.
This date is chosen for no particular reason
except to serve as a representative of all other days in our sample period;
the front month futures on this date is still quite some time 
before its expiration in September. 

It is evident that when the order flow is higher than the ``long-term'' mean, 
it tends to become smaller subsequently,
and vice versa.
This is a tell-tale sign that order flows are most likely mean reverting.
This empirical observation prompts us to assume the mean reversion process
in our mathematical modeling,
as in \eqref{e:X_t}.

Next, we observe in Figure \ref{fig:Order Flows} that 
JPX's NO futures in Panel B has the largest range of fluctuation,
which is about $\pm 3,000$ contracts.
JPX's NK futures has a range of about $\pm 400$ contracts.
By contrast, CME's yen-denominated NH futures fluctuates 
in the range of $\pm 150$ contracts,
and the dollar-denominated NX futures has a range of only $\pm 60$ contracts.

These observations are consistent with the descriptive statistics
captured in Table \ref{t:tickers}.
\[
\hbox{Figure \ref{fig:Order Flows} is about here.}
\]

\subsection{Estimation}

To estimate $\ell$, $p$, and $q$, 
we start from \eqref{e:d logP_t}.
Since 
\be
\dr \log S_t = \left(\mu_S-\half\sigma^2_S\right)\dr t + \sigma_S\, \dr z_t
\ee
from \eqref{e: dSt over St},
we consider the logarithmic return of $P_t$.
When discretized, we obtain
\be
r_t :=  \log P_t - \log P_{t-1}  
=  \left( \mu_S - \half \sigma_S^2 \right) \Delta t +  
f(X_t) - f(X_{t-1}) + \epsilon_t,
\label{e:econometric spec}
\ee
where $\epsilon_t$ is the noise term.
For sampling with fixed time interval,  
$\Delta t$ is a constant.
In particular,  $\Delta t = 1$ 
corresponds to 1-minute sampling period.
In this way, we obtain an econometric specification:
\be
r_t = a + f(x_t) - f(x_{t-1}) + \epsilon_t.
\label{e:regression}
\ee

By comparing against \eqref{e:econometric spec},
we can identify the intercept $a$ as
\[
a = \mu_S - \half \sigma_S^2.
\]
Comprising of $\mu_S$ and $\sigma_S$,
notice that the drift rate $a$ is purely a parameter
for the true price process.
For our modeling to be correct, the estimates for $a$
should be compatible for the 4 futures contracts.

The 1-minute time-series econometric specification \eqref{e:econometric spec}
is common to three price impact functions:
\begin{enumerate}
\item Heuristic square root function: $f(x) = \alpha\, \mathrm{sign}(x) \sqrt{|x|}$,
where $\mathrm{sign}(x)$ is either 1 for buyer-initiated trades,
$-1$ for seller-initiated trades, and 0 otherwise.
By definition, this heuristic function has an inflection point at $x=0$.
It is concave for $x>0$, and convex when $x <0$.
It is similar to the price impact function estimated by \cite{Hasbrouck:2004}.
\vspace{-0.5em}
\item Linear function \eqref{Equation: Linear Price Impact Function} 
, i.e., $f(x) = \alpha x$
\vspace{-0.5em}
\item S-shape function  \eqref{Equation: S-shape Function}
\end{enumerate}

For the heuristic square-root and linear functions,
we can use the ordinary least squares to estimate the respectrive $\alpha$'s.
As for the S-shape function,
we perform nonlinear regression \eqref{e:regression} under three constraints, 
as discussed earlier.
They are $\ell > 0$, $q > 0$, and \eqref{eq3.10}.
We put in place a grid of initial conditions for $p$ and $q$
to ensure that the estimates are robustly obtained.
For the 4 futures contracts on the same underlying Nikkei 225 index,
we present their respective parameter estimates in 
Table \ref{t:Estimation Results} for the regular trading session
on 15 June 2016 during the Asian hours, as an example.
\[
\hbox{Table \ref{t:Estimation Results} is about here.}
\]

\begin{figure}[th]
\caption{\small Price Impact Functions 
of Four Futures on 15 June 2016.}
\bigskip
\includegraphics[width=0.5\textwidth]{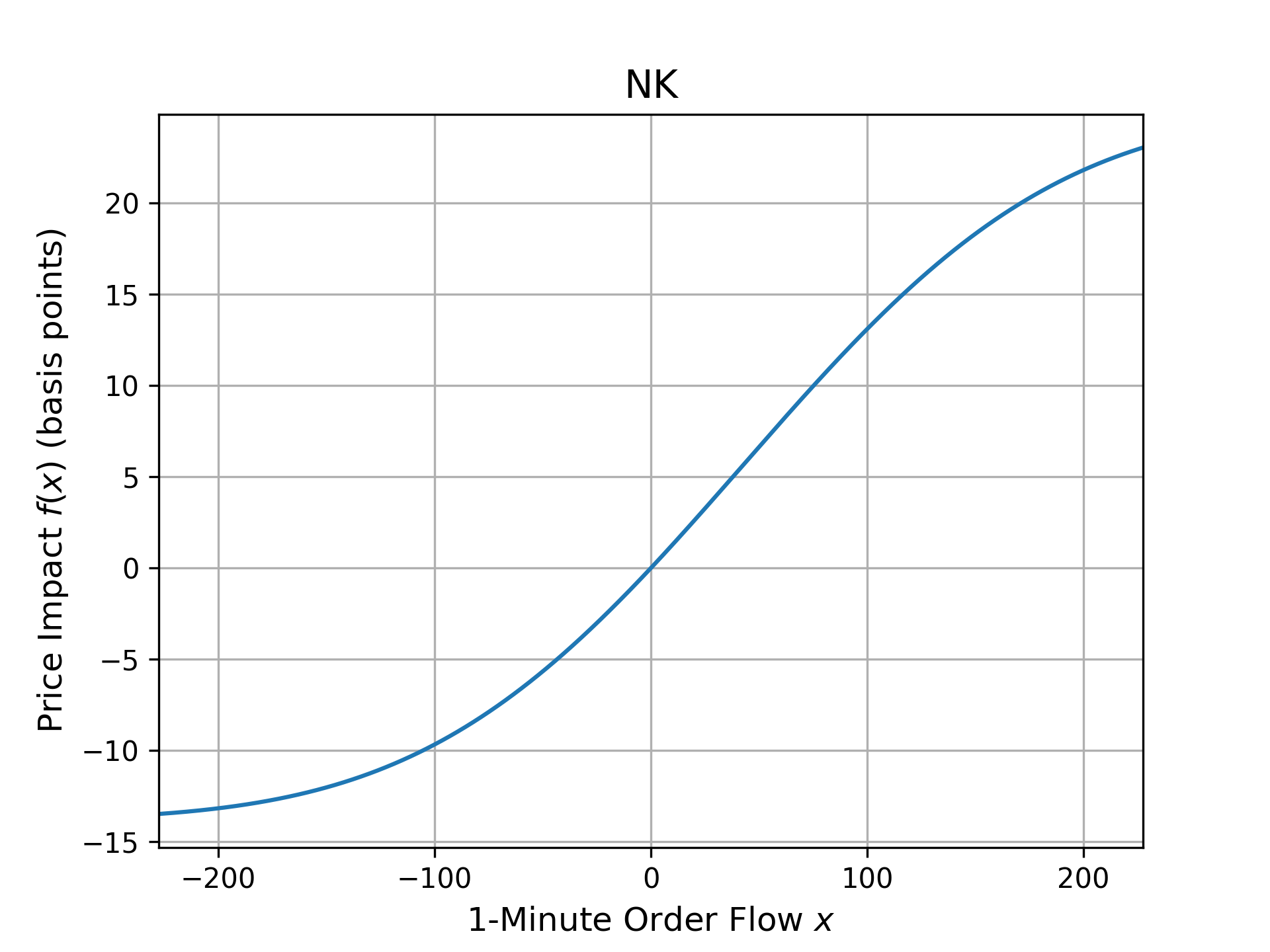}
\includegraphics[width=0.5\textwidth]{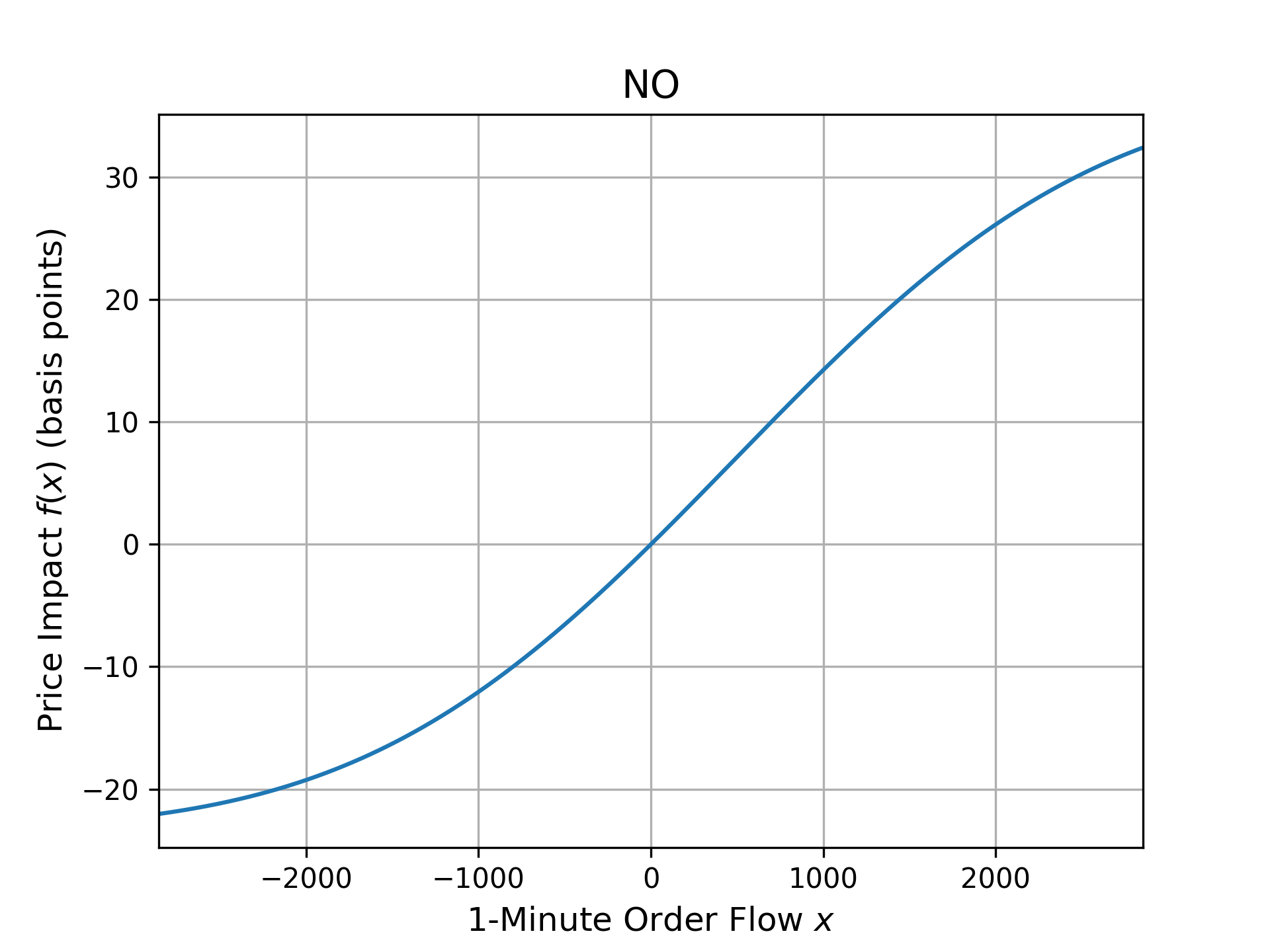}
\vskip3em
\includegraphics[width=0.5\textwidth]{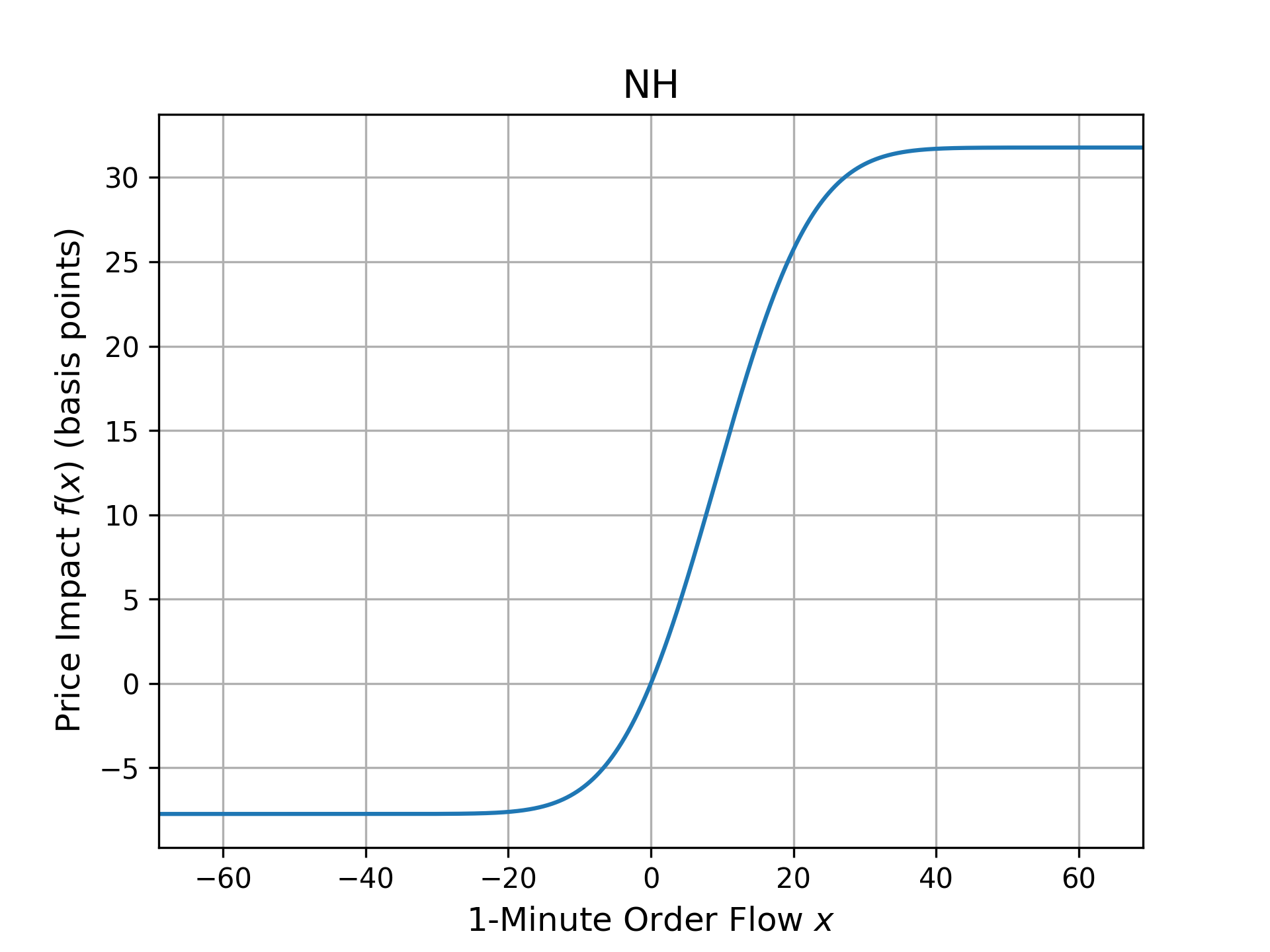}
\includegraphics[width=0.5\textwidth]{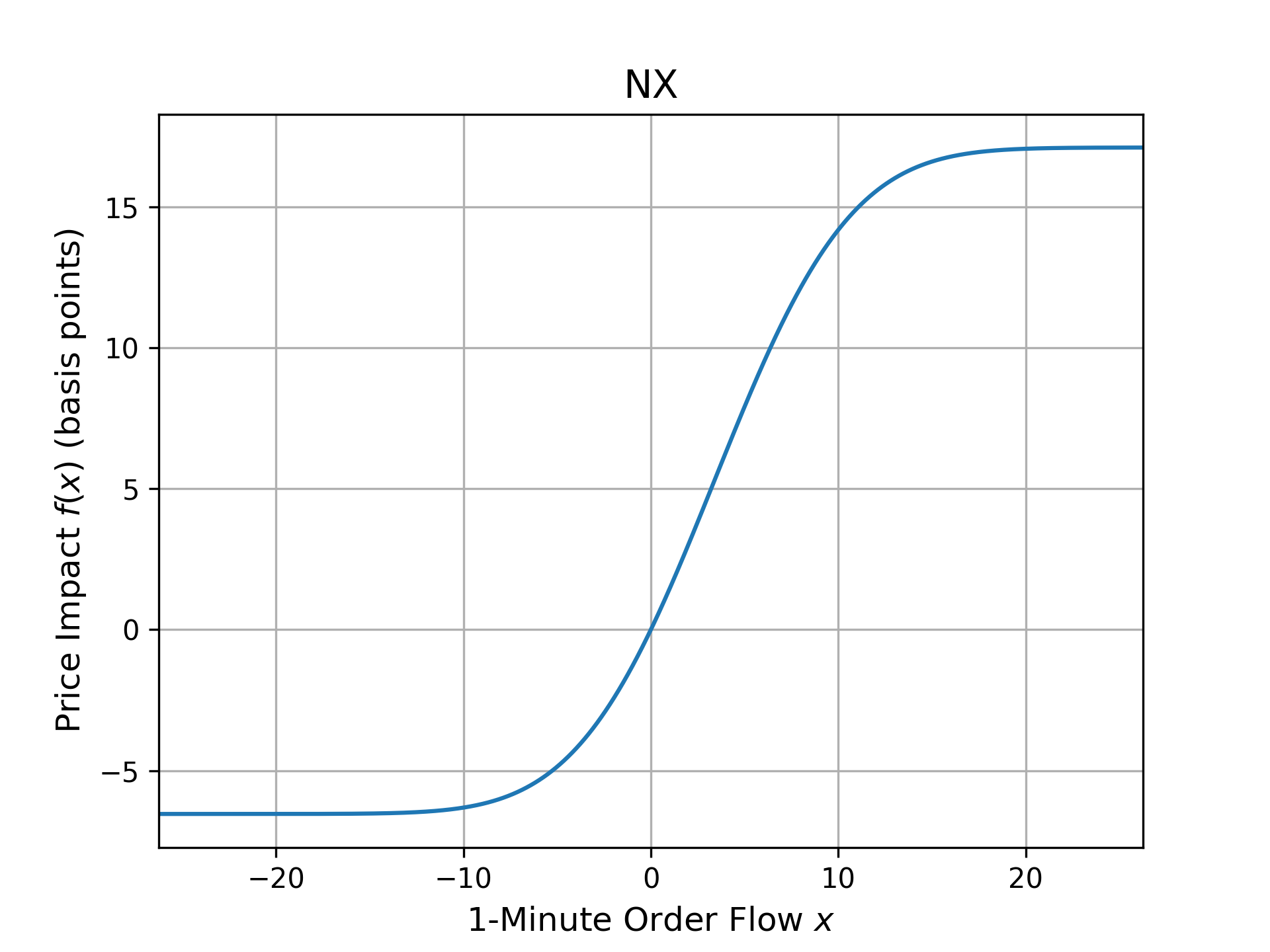}
\label{fig:impact functions}
\vfill
\end{figure}

Figure \ref{fig:impact functions} shows the 4 impact functions plotted
with the respective sets of estimated parameters, i.e.,
$\widehat{\ell}, \widehat{p}$, and $\widehat{q}$.
Clearly, we see that they are all S-shape curves.
It is interesting to note the asymmetry with respect to $x=0$.
When $\widehat{p}$ is estimated to be positive, 
negative order flow has a larger impact than positive order flow
of the same trade size, and vice versa.
This asymmetry is captured by Proposition \ref{s(x)=0}
as a result of the concave and convex nature of our $f(x)$ with respect to
an inflection point, which is not at $x=0$.
From Table  \ref{t:Estimation Results}, 
we see that all the $p$ estimates are negative,
and in Figure \ref{fig:impact functions}, 
it is evident that the price impact of $x >0$ is asymmetrically larger
on that particular day.

This asymmetry of buys versus sells could be due to the fact that the futures price
was moving higher on that day.
In line with this fact, Table \ref{t:Estimation Results} 
shows that all the drift estimates---intrinsic to 
the true price primarily---are positive.
It must be that there were more aggressive buy orders
willing to pay for the price impact cost,
pushing the price higher as a result. 
\[
\hbox{Figure \ref{fig:impact functions} is about here.}
\]

\begin{figure}[t]
\caption{\small Price Impact Functions 
of Four Futures Estimated over the Entire Sample Period.}
\includegraphics[width=0.5\textwidth]{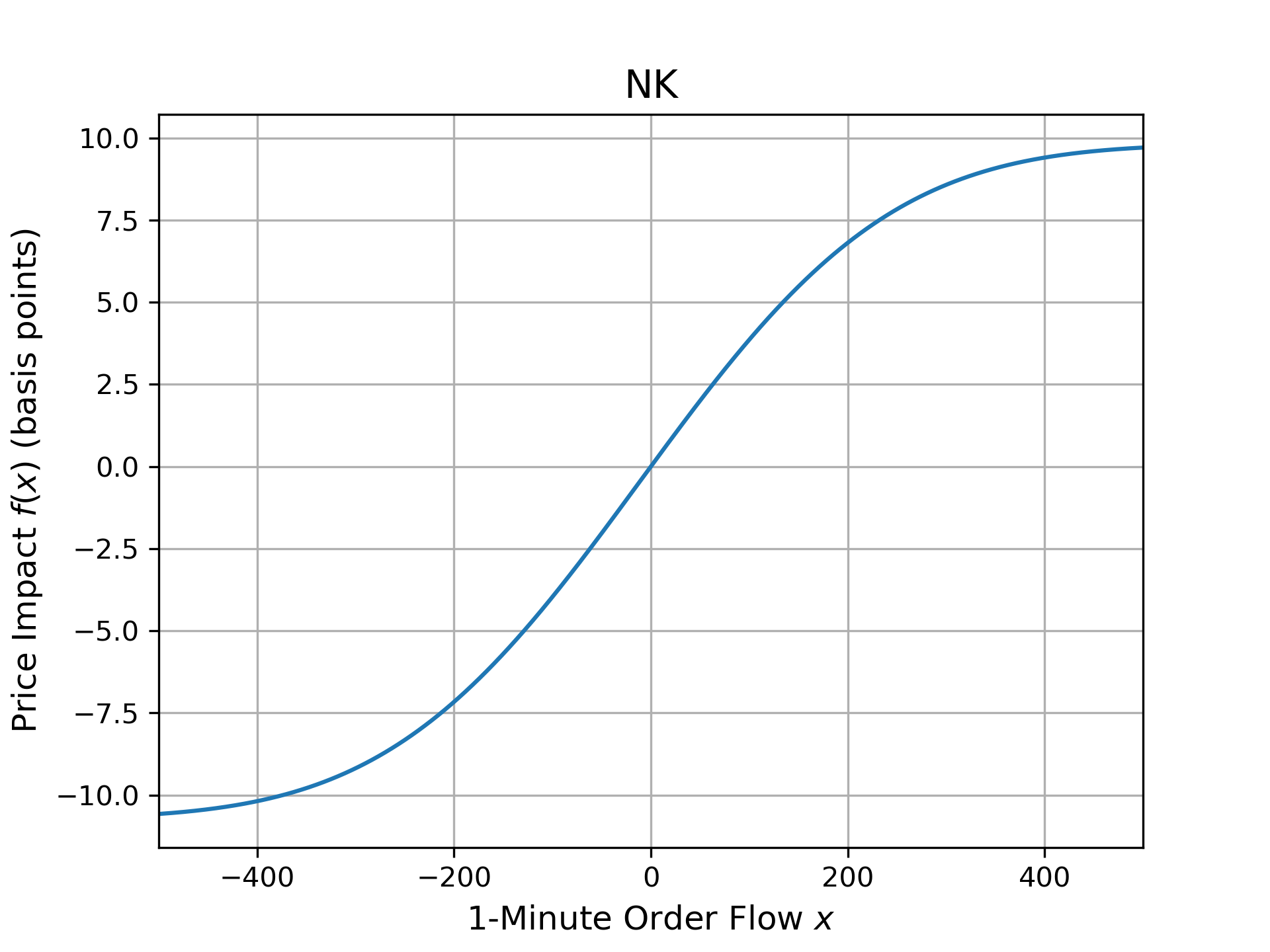}
\includegraphics[width=0.5\textwidth]{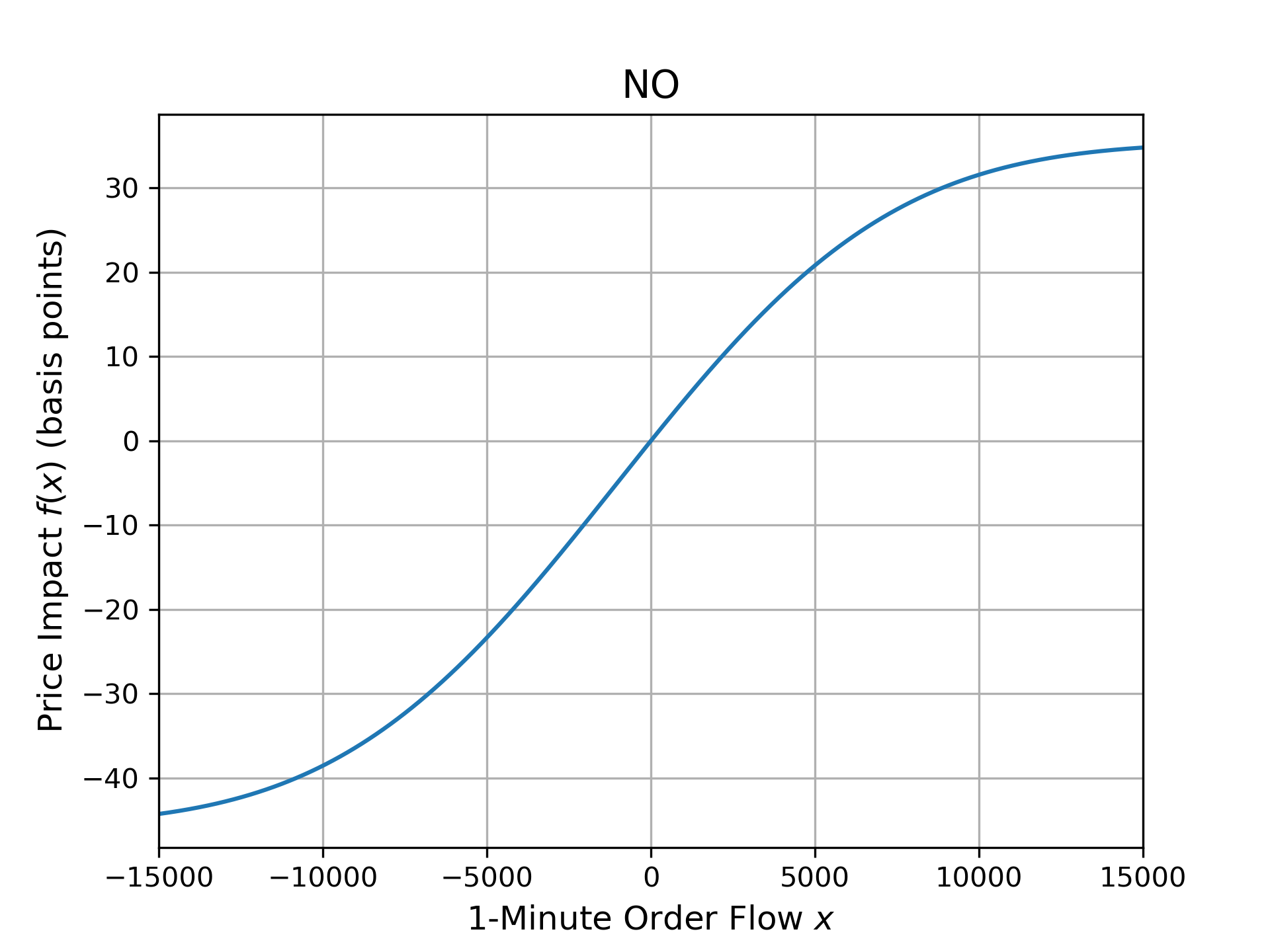}
\vskip3em
\includegraphics[width=0.5\textwidth]{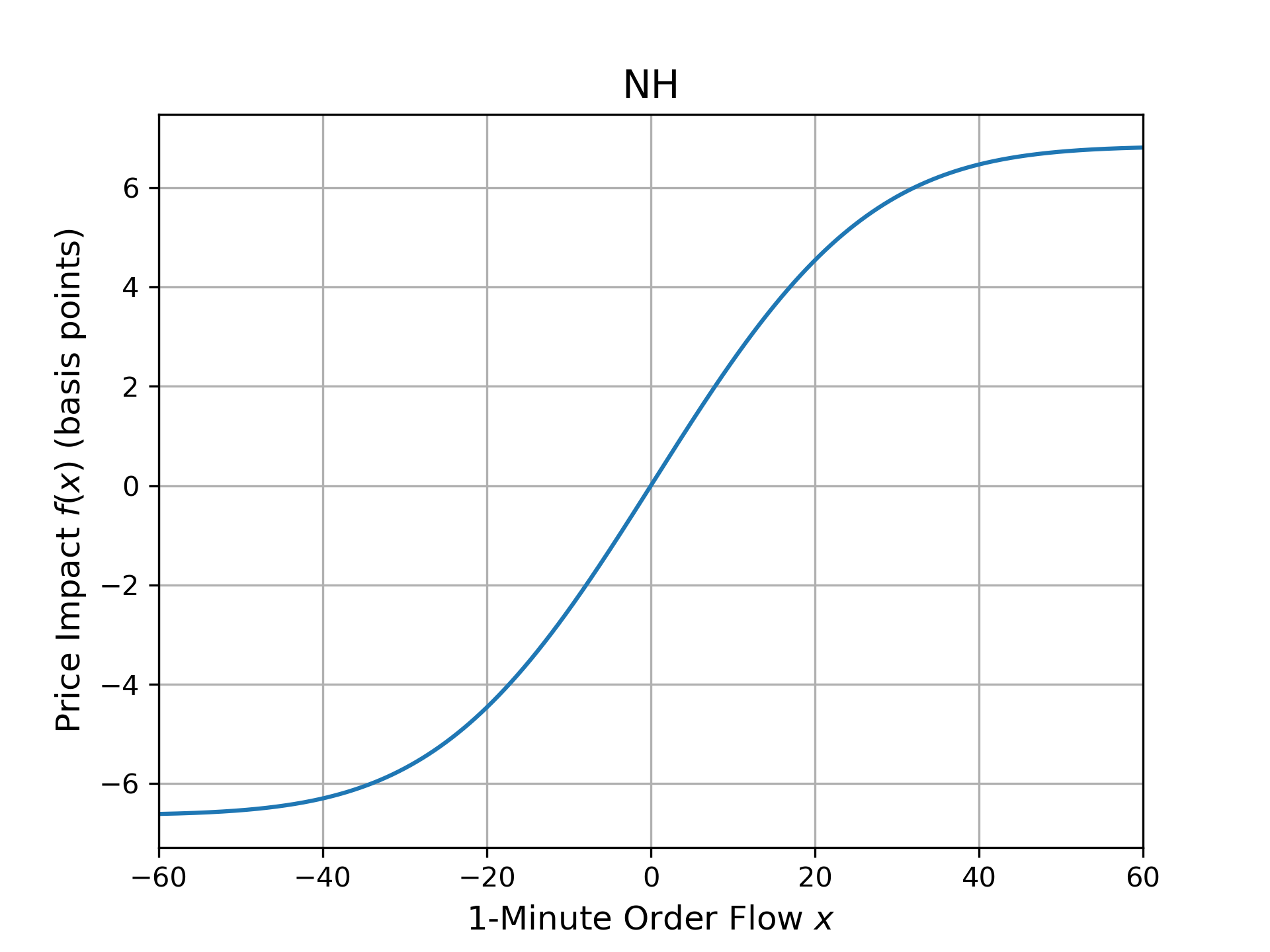}
\includegraphics[width=0.5\textwidth]{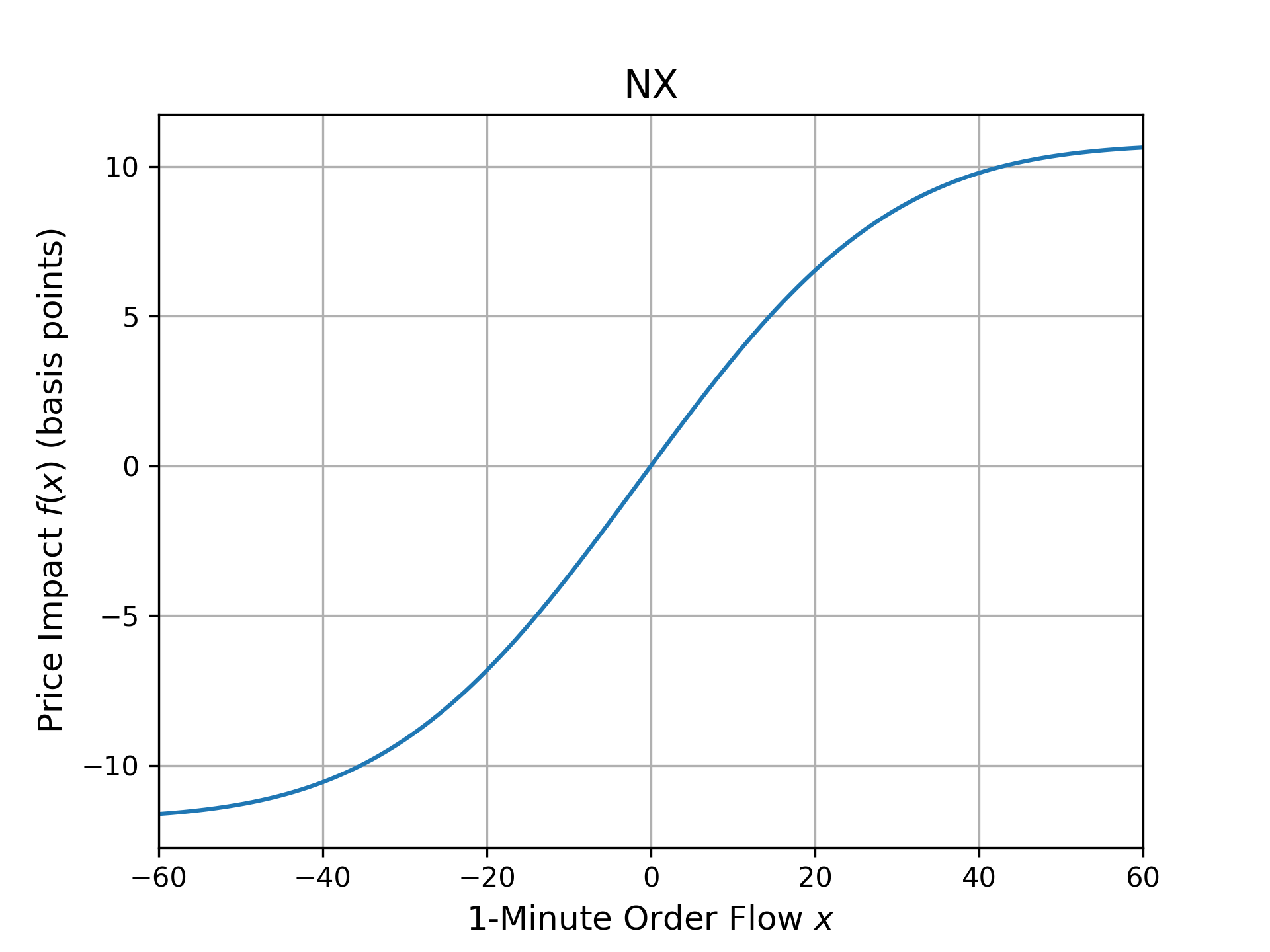}
\label{fig:impact functions all}
\vfill
\end{figure}

The asymmetry in the S-shape price impact function 
may occur on a given single day as shown in Figure~\ref{fig:impact functions}.
But would the asymmetry persist if we consider all the trading days 
over the entire sample period?
To answer this question, we pool together all the order flows
and the corresponding logarithmic returns,
and run the same constrained nonlinear regression.
The overall looks of the four S-shape functions
estimated respectively with pooled data from the entire sample period are displayed 
in Figure~\ref{fig:impact functions all}.
\[
\hbox{Figure \ref{fig:impact functions all} is about here.}
\]

We find that the S-shape impact functions for offshore NH and NX futures
in Figure~\ref{fig:impact functions all}
are much more symmetric compared to the corresponding pairs
in Figure~\ref{fig:impact functions}.
The price impact functions have a range of 
about $\pm$6 basis points for NH futures and $\pm$10 basis points for NX futures.
The onshore big Nikkei NK in Figure~\ref{fig:impact functions all} 
is  more symmetric too,
with the range of price impacts being approximately $\pm$10 basis points
for the order flows of $\pm$400 contracts.
To a lesser extent has the mini Nikkei NO in Figure~\ref{fig:impact functions all} 
become symmetric,
with buys being about 30 basis points for a large buy order flow of 10,000 contracts
and about 40 basis points for a large sell order flow of the same size.
For its single-day counterpart in Figure~\ref{fig:impact functions}, 
the difference of 10 basis points already starts to appear at only $\pm$3,000 contracts.

\subsection{Discussion of Estimation Results}
Recall that the parameter $\ell$ in the S-shape function
can be regarded as Kyle's illiquidity measure $\lambda$.
It is the price impact cost in basis points per unit of $\Phi(x)$
as in \eqref{e:Kyle}, i.e., when $\Phi(x)=1$. 
Looking at the $\ell$ estimates in Table \ref{t:Estimation Results},
it is clear that JPX's mini futures (NO) has the least value with only 0.01 basis points.
This outcome intuitively makes sense because mini futures
is the most heavily traded among the 4 contracts.
Overall, we observe that the two futures of JPX
are more liquid than CME's  offshore futures during the Asian hours of trading,
as anticipated.

Next, notice that in Table \ref{t:Estimation Results}, 
the order of magnitude for the $p$ and $q$ across different Nikkei 225 futures
is different.
In particular, $p$ and $q$ are the smallest for JPX's mini futures.
By contrast, $\widehat{a}$, 
which is exclusively about the unobservable trice price $S_t$
since $\ds a =  \mu_S -  \sigma_S^2/2$,
is more or less equal across the 4 futures.
This result is consistent with the fact that these 4 futures are written on 
the same underlying equity index,
and therefore their unobservable prices $S_t$ should move in tandem
generally with the underlying index.
This result provides evidence that our modeling and econometrics
are on the right track.
\[
\hbox{Table \ref{t:Estimation Results all} is about here.}
\]

Now, in Table~\ref{t:Estimation Results all},
we report the estimation results for each of the 4 futures contracts
when the order flows and the corresponding log returns 
in the entire sample period of about 3 years and 7 months are pooled together.
Their overall  S-shape curves have already been plotted 
in Figure~\ref{fig:impact functions all}.

First, we find that the parameter estimates in Table~\ref{t:Estimation Results all}
are smaller than the corresponding estimates in Table~\ref{t:Estimation Results},
by one to two orders of magnitude.
This finding is to be expected because
when one considers intra-day data over many days,
one will find that the overall negative order flows 
and the overall positive order flows tend to counteract each other.
If for a certain intra-day time period in a trading day,
the order flow is a big positive
and for another time period in another trading day,
the order flow is a big negative,
then there is a higher likelihood that they both occur and thus
balance out each other in the nonlinear regression.

Second, despite the smaller parameter estimates,
their $t$ statistics are however larger by comparison
with Table \ref{t:Estimation Results} in general.
The main exception is the $t$ statistics for the drift $\widehat{a}$,
which are compatible with the single-day $t$ statistics.
The adjusted $R^2$ are lower in Table~\ref{t:Estimation Results all}.
This is because the number of pairs of the order flow and the log return
has become a lot larger (about a thousand times) in the pooled regression.
Extreme large order flows and log returns are more likely to occur,
which tend to make the residual sum of squares larger,
resulting in a lower adjusted $R^2$.

Interestingly, the findings from the single-day results
are also observed in multi-day Table~\ref{t:Estimation Results all}.
Again, we find that $\widehat{a}$'s estimates (after multiplied by $10^6$) 
of 1.04, 1.05, 1.01, and 1.00  are numerically compatible across
the 4 futures contracts.
As mentioned earlier, this compatibility must occur
because these 4 futures contracts are written on the same underlying Nikkei 225 index,
and they should move in the same direction at almost the same drift rate.
As anticipated, we find that NO futures contract is the most liquid
with the smallest $\widehat{\ell}$, $\widehat{p}$, and $\widehat{q}$ estimates.
The offshore NX futures contract, which is denominated in US dollars,
is the least liquid overall during the Asian trading hours.

In summary, the findings from Tables~\ref{t:Estimation Results}
and \ref{t:Estimation Results all},
are consistent with the stylized facts 
about these 4 Nikkei 225 index futures traded concurrently
during the Asian hours.
As anticipated, the mini futures is most liquid,
and CME's futures are comparatively less liquid.
Moreover, the drift rate $a$, being primarily about the spot Nikkei 225 index
that is common to and shared by these 4 futures, 
is more or less equal statistically,
despite the fact that their liquidity levels differ significantly.

\subsection{Comparisons}
Now we present a comparison of the heuristic function
against our linear and S-shape functions
in capturing the price impact.
Table \ref{t:Comparison} displays the adjusted $R^2$
to examine the goodness of fit, and the Bayesian information criterion (BIC)
to evaluate over-fitting vis-\`{a}-vis under-fitting. 

A point to note is that for the S-shape function,
as discussed previously, as in \eqref{e:Kyle},
the parameter $\ell$ plays a role that is analogous to the $\alpha$ parameter 
in the heuristic and linear functions.

Again, we note that the drift parameter $a$ has 
quantitatively similar estimate $\widehat{a}$, 
which is not statistically significant at the 5\% level.
By contrast, the $\alpha$ estimates are all statistically significant,
and their values are rather different.
In particular, the $\alpha$ estimate is the smallest for our linear function,
being of an order of magnitude smaller than the corresponding estimates
for the heuristic and the S-shape function.
Overall, as expected, JPX's NO futures has the smallest $\widehat{\alpha}$,
and CME's NX futures has the largest  $\widehat{\alpha}$
during the Asian trading hours.

More importantly, we find that the adjusted $R^2$ for the S-shape function
is higher than those of linear and heuristic functions.
Also the residual sum of squares is the least by comparison.
Therefore, we have evidence that the S-shape function fits the data better.
From the Bayesian information criterion (BIC),
we see that the 3-parameter S-shape function has s much lower BIC 
compared to those of 1-parameter heuristic and linear functions,
which means that over-fitting by 3 parameters did not occur.
\[
\hbox{Table  \ref{t:Comparison}  is about here.}
\]

\subsection{All Estimations}

Now, we examine whether the evidence from a single day (15 June 2016) 
presented earlier does hold for the entire set of estimates.
From the beginning of November 2013 through early June 2017 for a total of 885 days
for JPX's onshore futures and 927 days for CME's offshore futures\footnote{
Being offshore, 
CME's futures are electronically traded even when JPX has no trading because of holidays.
Japan has more public holidays than the United States.},
we obtain by \eqref{e:econometric spec} 
a set of $\ell, p$ and $q$ estimates and other statistics 
from the 1-minute intra-day data for each day.

We run the paired sample tests
of their adjusted $R^2$ for each trading day in our sample period.
The paired $t$ test results in Panel A of Table \ref{t:test} 
shows that the heuristic square-root function
in general has a better fit with the data than the linear function,
with the exception of JPX's mini futures contract (NO).
On average, the adjusted $R^2$ of the linear function is about 3.16 percentage points
lower than that of the heuristic function in the case of CME's NH futures,
and the $t$ statistic of -28.53 suggests that this difference is highly significant.
In a way, it could be the reason why the industry uses the square-root function
rather than the linear function in their analysis of price impact.

Panel B of Table \ref{t:test} shows the $t$ test results for 
comparing the residual sum of squares (RSS). 
For NK and NH futures contracts, the $t$ statistics are significant 
and we can conclude that for our samples,
the linear function has larger errors in fitting the data 
compared to the heuristic function.
For NX futures, however, the difference in RSS is not significant.
On the other hand,
for NO futures, our linear function's RSS is significantly smaller
than that of the heuristic function.

\clearpage

More importantly, the $t$ test statistics for
the comparison between our S-shape function and the heuristic function
indicate that for all 4 futures contracts, 
the differences in adjusted $R^2$ and RSS are statistically significant.
For the onshore NK and NO contracts, the S-shape function is better than
the heuristic function by 2.08 and 2.92 percentage points, respectively.
For the off-shore NH and NX contracts, the difference is about 1.5 percentage points.
The RSS incurred by our S-shape function is much smaller.
Also, given that the BIC is on average smaller than the BIC 
for the single-parameter heuristic function by 1,000,
our 3-parameter S-shape function does not have the problem of over-fitting.
The empirical evidence points to the finding that our S-shape price impact function
is more capable in reflecting the price impact of a Nikkei futures trade.
\[
\hbox{Table \ref{t:test} is about here.}
\]

Next, we compare the liquidity 
of the 4 futures contracts on the same underlying equity index.
For this purpose, in Table \ref{t:ds}, we provide the descriptive statistics 
for the parameter estimates of the S-shape functions
over the sample period.
The average and standard deviation, along with the percentiles, are tabulated.
We note that the distributions of the parameter estimates are skewed to the right,
so much so that the average and the 50-th percentile (median) are quite different.

For the $\ell$ parameter estimates, which mirrors the role of
$\lambda$ parameter in \cite{Kyle:1985},
we find that, as anticipated, the NO futures is the most liquid
as its average of 0.10 basis points (bps) is the smallest. 
Moreover, for each percentile in Table \ref{t:ds}, 
NO futures' $\ell$ estimate is the smallest among the 4 futures contract specifications.
As shown in \eqref{e:Kyle}, a small $\ell$ leads to a small disparity or spread
from the true price, and vice versa.

By contrast, based on the percentiles, the dollar-denominated NX futures
is the least liquid during Asian trading hours,
since its $\ell$ estimates are the largest for each percentile in Table \ref{t:ds}
up to the median value of 0.49 basis points.
Beyond the 90-th percentile, JPX's big Nikkei futures NK is the least liquid.
This empirical finding is consistent with the fact
that the minimum tick size of NK futures is twice that of 
all the other 3 futures contracts.
Furthermore, the contract size of NK futures 
is 10 times that of NO futures and twice that of NH futures.
It is a stylized fact that in market microstructure,
a larger tick size and contract size will lead to a larger quoted spread,
everything else being equal.
\[
\hbox{Table \ref{t:ds} is about here.}
\]

Finally, it is interesting to note that despite the differences in the 
average $\ell$, $p$, and $q$ estimates,
the average inflection points in Table~\ref{t:ds}
are compatible with the estimates obtained from pooled observations
in Table~\ref{t:Estimation Results all}.

\subsection{Market Depth}
The inflection point, defined in \eqref{Equation: Inflection Point}
as primarily a ratio of $p$ and $q$,
may be interpreted as the market depth in relation to the market price movement.
By construction, the inflection point is in the unit of order flow (number of contracts).
For our S-shape function, the unique inflection point is 
the order flow at which the gradient of the function is at its maximum,
which means that the price impact from the order flow is at its highest.
Thus, if the inflection point is far away from the origin,
it means that a larger volume of order flow is needed to create the maximum impact.
Put differently, the market is deep enough to ameliorate
the price impact from small- to medium-sized order flows.

Interestingly, according to \eqref{Equation: Inflection Point 2},
the inflection point is dependent on 
2 parameters of the market price of liquidity risk ($\delta$ and $\tau$),
3 quantities of the order flow ($c$, $m$, and $\eta$), 
and the volatility $\sigma_S$ of the true price return
through its correlation $\rho$  with the order flow.
As a matter of fact, almost all the parameters in our framework
appear in the calculation of the inflection point.

As anticipated, the mini futures NO with the smallest contract size
has the largest inflection point in absolute value.
In Table \ref{t:ds}, NO futures has an average inflection point of -1,654 contracts.
Coming in second is NK futures with an average of 28 contracts.
Obviously during the Asian trading hours, JPX's onshore futures are more active
and the market depth is relatively deeper than the offshore futures 
with an average of 3 contracts only.
\[
\hbox{Table \ref{t:dsbest} is about here.}
\]

Now, it is well known that market depth is intrinsically related to
the amount of liquidity in the (electronic) market, i.e.,
the numbers of contracts waiting to buy and to sell.
Our data sets contain (only) the best bid and best offer prices 
along with their respective sizes.
In Table \ref{t:dsbest}, we present the descriptive statistics
of the best bid and ask sizes at the beginning of each 1-minute interval.
It is evident that NO futures have the largest bid and ask sizes of 332
and 341 contracts on average, respectively,
over our sample period of about 3 years and 7 months.
The big Nikkei NK Futures comes in second.
Understandably, during the Asian trading hours, 
the offshore NH and NX futures of CME have rather thin liquidity of
less than 10 contracts.
The statistics in \ref{t:dsbest} 
provide collaborative evidence based on futures' best quotes.
They support the idea that the inflection point of the S-shape function
may be conceived as a measure for market depth.

\section{Conclusions} \label{Conclusions}
Starting from the framework in \cite{CJP:2004}, 
we introduce the key proposition
that the order flow is a mean reversion process.
Applying It\^{o}'s calculus, and with a change of probability measure
involving the market price of market risk
and the market price of liquidity risk, 
we obtain an inhomogeneous Bernoulli differential equation.
By solving the equation in two special cases,
we obtain two closed-form solutions for the functional form
of the price impact function.
The first case yields a linear impact function that is consistent with \cite{Kyle:1985},
and the second case produces a nonlinear S-shape function.

Within our mathematical framework, 
we explore the theoretical implications for a possible explanation 
of the phenomenon that the volatility is stochastic.
In principle, order flow may drive the price temporarily
above or below the unobservable fundamental value.
In other words, 
order flow as a manifestation of trading will give rise to additional volatility
that is otherwise absent.
This excess volatility is stochastic because the order flow is stochastic.

Our theoretical paper also includes an empirical study.
Using 4 different futures contracts of different specifications 
but on the same underlying Nikkei 225 index,
we find that the estimates from constrained nonlinear least squares method
are consistent with the stylized facts about the liquidity of these futures.
We also provide evidence indicative of the superiority of our S-shape function
over the heuristic square-root functions. 
Interestingly, our empirical analysis shows that the inflection point
of the S-shape function can be interpreted as the market depth.

Market participants and researchers 
are interested in estimating the market depth, 
and the financial implications of price impact.
As a possible application of the S-shape price impact function,
we show that the notion of market depth can
be implemented to capture the level of liquidity risk in trading,
much like volatility is a measure of price uncertainty or risk.
Another application is to use the price impact function
for pre-trade and post-trade analyses 
to estimate the price impact cost of trading a certain amount
of financial instrument.
Although our empirical study focuses on Nikkei 225 futures for checking
the correctness of our model,
our framework is general and potentially applicable to 
futures on commodity and on interest rates, forex, as well as stocks.


\section*{Acknowledgments}
We would like to sincerely thank an anonymous reviewer for
providing many insightful comments and suggestions,
which improve the quality of the paper considerably.
One of the authors (Kijima) is grateful for the research grants 
funded by the Grant-in-Aid (A) ({\#}26242028) 
from Japan's Ministry of Education, Culture, Sports, Science and Technology.

\clearpage




\clearpage
\begin{table}[htbp]
\begin{center}
{\small
\centering
\caption{Descriptive Statistics of Nikkei 225 Index Futures Contracts}
\label{t:tickers}
\bigskip
\begin{tabular}{lcccrrrrr}
\toprule
&     &     &     & \multicolumn{5}{p{20.325em}}
{\hfil Daily Average (Contracts)\hfill} \\
\cmidrule{5-9}\multicolumn{1}{p{4.565em}}{\hfil Name\hfill} &
\multicolumn{1}{p{4.75em}}{Bloomberg} & 
\multicolumn{1}{p{3.59em}}{Contract} & 
\multicolumn{1}{p{1.94em}}{Tick} &
\multicolumn{1}{p{3.315em}}{Volume} & 
\multicolumn{1}{p{3.0em}}{Positive} &
\multicolumn{1}{p{3.5em}}{Negative} &
\multicolumn{1}{p{2.315em}}{Order} &
\multicolumn{1}{p{5.5em}}{Volatility of} \\
\multicolumn{1}{p{4.565em}}{of Futures} &  
\multicolumn{1}{p{4.75em}}{\hfil Symbol\hfill}   & 
\multicolumn{1}{p{3.59em}}{\hfil Size\hfill} & 
\multicolumn{1}{p{1.94em}}{\hfil Size\hfill} &     & 
\multicolumn{1}{p{3.0em}}{\hfil OF\hfill} & 
\multicolumn{1}{p{3.5em}}{\hfil OF\hfill} &
\multicolumn{1}{p{2.315em}}{Flow (OF)} & 
\multicolumn{1}{p{2.5em}}{OF\hbox{\hspace{1.0em}$\big(\widehat{\eta}\big)$}} \\
\midrule
JPX Big Nikkei  &  NK     &  \yen 1,000  & 10   & 52,368 & 14,013 & 13,572 & 1.23 & 162.03 \\
JPX Mini Nikkei  &  NO     &  \yen 100    & 5    & 441,585 & 75,377 & 75,885 & -1.41 & 700.79 \\
\hdashline
CME Nikkei  &  NH     &  \yen 500      & 5    & 7,331 & 1,358 & 1,420 & -0.17 & 17.22 \\
CME Nikkei  &  NX     &  \$50    & 5    & 3,702 & 742  & 768  & -0.07 & 8.62 \\
\hline
\bottomrule
\end{tabular}%
}
\end{center}
\end{table}
\vskip-1.5em
\noindent
{\small
This table tabulates the 4 futures contracts 
on the same underlying Nikkei 225 index for 9 AM to 3 PM
Japan Standard Time.
The sample period is from November 2013 through June 2017, about 3 years and 7 months.
The column labeled as ``Order Flow (OF)'' is the average of 1-minute order flow.
Daily average order flow can be inferred from the difference
between the positive and negative order flows. 
The column labeled as ``Volatility of OF $\big(\widehat{\eta}\big)$''
contains the volatility estimates for the 1-minute order flows averaged
over the sample period.
}

\clearpage

\begin{table}[htbp]
\centering
\caption{Estimation Results of S-Shape Function of Order Flow on 15 June 2016}
\vskip2em
\begin{tabular}{lrrrr}
\toprule
& \multicolumn{1}{c}{NK} & \multicolumn{1}{c}{NO} 
& \multicolumn{1}{r}{NH} & \multicolumn{1}{r}{NX} \\
\midrule
$\widehat{a}\times 10^5$ & 1.97 & 2.14 & 2.34 & 2.11 \\
$t\big(\widehat{a}\big)$ & 0.59 & 0.78 & 0.77 & 0.69 \\
\hdashline
$\widehat{\ell}$ (in bps) & 0.13 & 0.01 & 0.97 & 1.39 \\
$t\big(\widehat{\ell}\big)$  & 4.63 & 5.52 & 2.76 & 2.63 \\
\hdashline
$\widehat{p}$ (\%) & -0.34 & -0.02 & -7.06 & -10.49 \\
$t\big(\widehat{p}\big)$ & -1.68 & -1.36 & -1.71 & -1.39 \\
\hdashline
$\widehat{q}\times 10^5$  & 8.15 & 0.04 & 653.35 & 3,076.31 \\
$t\big(\widehat{q}\big)$ & 2.03 & 1.76 & 1.55 & 1.39 \\
\hdashline
Inflection (Market Depth) & 42 & 478 & 11 & 3 \\
\hdashline
Adjusted $R^2$ (\%) & 27.84 & 38.98 & 18.40 & 16.09 \\
\bottomrule
\end{tabular}%
\label{t:Estimation Results}%
\end{table}%
\noindent
{\small
One-minute order flows are obtained for each of the 4 futures contracts
over the trading session (9 AM to 3 PM Japan Standard Time) on 15 June 2016.
These 4 futures are written on the same underlying Nikkei 225 Index.
Through a constrained nonlinear least squares procedure,
this table presents the parameter estimates 
$\big(\widehat{p}, \widehat{q}$, and $\widehat{\ell}\,\big)$
of our S-shape price impact function, 
together with the estimate $\widehat{a}$
for the drift rate intrinsic to the true price. 
The $t$ statistic is shown beneath each estimate.
}

\clearpage

\begin{table}[htbp]
\centering
\caption{
Estimation Results of S-Shape Function of Order Flow for the Entire Sample Period
}
\label{t:Estimation Results all}%
\vskip2em
\begin{tabular}{lrrrr}
\toprule
& \multicolumn{1}{c}{NK} & \multicolumn{1}{c}{NO} 
& \multicolumn{1}{r}{NH} & \multicolumn{1}{r}{NX} \\
\midrule
$\widehat{a}\times 10^6$ & 1.04 & 1.05 & 1.01 & 1.00 \\
$t\big(\widehat{a}\big)$ & 0.59 & 0.88 & 0.87 & 0.88 \\
\hdashline
$\widehat{\ell}$ (in bps) & 0.041 & 0.005 & 0.261 & 0.372 \\
$t\big(\widehat{\ell}\big)$  & 59.60 & 57.48 & 55.06 & 63.99 \\
\hdashline
$\widehat{p}$ (in bps) & 2.73 & 0.24 & -9.21 & 23.67 \\
$t\big(\widehat{p}\big)$ & 3.49 & 6.73 & -1.15 & 3.47 \\
\hdashline
$\widehat{q}\times 10^6$  & 24.80 & 0.02 & 2,358.20 & 1,719.63 \\
$t\big(\widehat{q}\big)$ & 23.76 & 11.27 & 23.14 & 22.85 \\
\hdashline
Inflection (Market Depth) & -11 & -1,037 & 0 & -1 \\
\hdashline
Adjusted $R^2$ (\%) & 5.86 & 12.31 & 4.94 & 6.44 \\
\bottomrule
\end{tabular}%
\end{table}%
\noindent
{\small
One-minute order flows are obtained for each of the 4 futures contracts
over the trading session (9 AM to 3 PM Japan Standard Time)
for the entire sample period from early June 2013 through early July 2017.
These 4 futures are written on the same underlying Nikkei 225 Index.
Through a constrained nonlinear least squares procedure,
this table presents the parameter estimates 
$\big(\widehat{p}, \widehat{q}$, and $\widehat{\ell}\,\big)$
of our S-shape price impact function, 
together with the estimate $\widehat{a}$
for the drift rate intrinsic to the true price. 
The $t$ statistic is shown beneath each estimate.
}

\clearpage

\begin{table}[htbp]
\begin{center}
{\small
\centering
\caption{Comparison of Heuristic, Linear, and S-Shape Functions}
\label{t:Comparison}
\bigskip
\begin{tabular}{clcccrccr}
\toprule
  &      & $\widehat{a}\times 10^{5}$ & $t(\widehat{a})$ & 
$\widehat{\alpha}$ (in bps) & \multicolumn{1}{l}{$t(\widehat{\alpha})$} & 
Adj $R^2$ & \multicolumn{1}{l}{RSS$\times 10^{5}$} & BIC \\
\midrule
    & Heuristic & 1.97 & 0.58 & 0.3691 & 10.62 & 26.63\% & 14.37 & -4,258 \\
NK  & Linear & 2.02 & 0.59 & 0.0293 & 7.46 & 24.41\% & 14.80 & -4,247 \\
    & S-Shape & 1.97 & 0.59 & 0.1308 & 4.63 & 27.84\% & 14.06 & -5,273 \\
\hdashline
    & Heuristic & 2.13 & 0.71 & 0.1457 & 14.07 & 36.35\% & 10.08 & -4,385 \\
NO  & Linear & 2.14 & 0.71 & 0.0046 & 13.49 & 38.02\% & 9.82 & -4,395 \\
    & S-Shape & 2.14 & 0.78 & 0.0139 & 5.52 & 38.98\% & 9.62 & -5,409 \\
\midrule
    & Heuristic & 2.26 & 0.69 & 0.5411 & 7.23 & 14.93\% & 12.17 & -4,318 \\
NH  & Linear & 2.27 & 0.69 & 0.0720 & 4.14 & 11.12\% & 12.72 & -4,302 \\
    & S-Shape & 2.34 & 0.77 & 0.9707 & 2.76 & 18.40\% & 11.62 & -5,341 \\
\hdashline
    & Heuristic & 2.12 & 0.67 & 0.7265 & 6.60 & 14.64\% & 12.05 & -4,321 \\
NX  & Linear & 2.10 & 0.66 & 0.1428 & 4.96 & 10.35\% & 12.66 & -4,303 \\
    & S-Shape & 2.11 & 0.69 & 1.3894 & 2.63 & 16.09\% & 11.80 & -5,336 \\
\bottomrule
\end{tabular}%
}
\end{center}
\end{table}
\vskip-1.5em
\noindent
{\small
This table tabulates a comparison of the heuristic square root, linear,
and S-shape functions of order flow in capturing the price impact
for the trading session on 15 June 2016 (9 AM to 3 PM Japan Standard Time).
The $\ell$ parameter of the S-shape function is used in this comparison
with the $\alpha$ parameter in the heuristic and linear functions.
$\widehat{a}$ is the estimate of the intercept
and $t(\widehat{a})$ is its $t$ statistic,
likewise for the \cite{Kyle:1985} illiquidity measure $\alpha$.
In addition to the adjusted $R^2$, 
we also include the residual sum of squares (RSS) to
measure the goodness of fit. 
The Bayesian information criterion (BIC) shows whether the
S-shape function has an over-fitting problem.
}

\clearpage

\begin{table}[htbp]
\centering
\caption{Two-Sample $t$ Tests of Goodness of Fit}
\label{t:test}
\bigskip
\begin{center}
Panel A: Adjusted $R^2$
\\[0.5em]
\begin{tabular}{l|cc|cc}
\hline
& \multicolumn{2}{c|}{Linear $-$ Heuristic} 
& \multicolumn{2}{c}{S-shape $-$ Heuristic} \\
& mean difference & $t$ statistic & mean difference & $t$ statistic \\
\hline
NK  & -2.34\% & -20.30 & 2.08\% & 15.91 \\
NO  & 1.44\% & 11.91 & 2.92\% & 26.75 \\
\hdashline
NH  & -3.16\% & -28.53 & 1.48\% & 17.27 \\
NX  & -1.60\% & -13.08 & 1.56\% & 17.48 \\
\hline
\end{tabular}%
\end{center}
\end{table}
\begin{center}
Panel B: Residual Sum of Squares
\end{center}
\begin{center}
\begin{tabular}{l|cc|cc}
\hline
& \multicolumn{2}{c|}{Linear $-$ Heuristic} 
& \multicolumn{2}{c}{S-shape $-$ Heuristic} \\
& mean difference $\times 10^6$ & $t$ statistic 
& mean difference $\times 10^6$ & $t$ statistic \\
\hline
NK  & 2.89 & 2.51 & -7.63 & -5.43 \\
NO  & -3.65 & -3.09 & -5.81 & -3.59 \\
\hdashline
NH  & 1.58 & 3.95 & -2.44 & -2.07 \\
NX  & 0.32 & 0.67 & -3.35 & -2.28 \\
\hline
\end{tabular}%
\end{center}
\vspace{1em}
\noindent
{\small
This table tabulates the results of two-sample $t$ tests.
Under the null hypotheses that the adjusted $R^2$ is no different in Panel A
and the residual sum of squares is no different in Panel B,
we test our linear and non-linear price impact functions against
the heuristic square root function.
This comparison is over the entire sample period from early November 2013
through early June  2017.
}

\clearpage

\begin{table}[htbp]
\centering
\caption{Descriptive Daily Statistics for S-Shape Price Impact Functions}
\vskip1em
\label{t:ds}
\begin{footnotesize}
\begin{tabular}{l@{}c@{\:\:}r@{\:\:}r@{\:\:}r@{\:\:}r@{\:\:}r@{\:\:}r@{\:\:}r@{\:\:}r@{\:\:}r@{\:\:}}
\toprule
     &      &      &      & \multicolumn{7}{c}{Percentile} \\
\cmidrule{5-11} {Estimate} & Symbol & {Average} & STD  
& 1    & 5    & 10   & 50   & 90   & 95   & 99 \\
\midrule
\multicolumn{1}{l}{$\ell$} & NK  & 5.17 & 16.83 & 3.3$\times10^{-5}$ & 0.01 & 0.02 & 0.12 & 12.78 & 20.20 & 84.73 \\
\multicolumn{1}{l}{(bps)} & NO  & 0.10 & 0.63 & 1.3$\times10^{-3}$ & 1.9$\times10^{-5}$ & 2.4$\times10^{-5}$ & 0.01 & 0.02 & 0.26 & 2.24 \\
    & NH  & 0.93 & 3.99 & 0.05 & 0.08 & 0.12 & 0.32 & 0.94 & 1.58 & 13.45 \\
    & NX  & 1.87 & 12.54 & 0.09 & 0.16 & 0.21 & 0.49 & 1.29 & 2.34 & 28.86 \\
\midrule
\multicolumn{1}{l}{$p$} & NK  & -0.28 & 4.07 & -25.98 & -2.48 & -0.83 & 1.0$\times10^{-4}$ & 0.66 & 2.11 & 8.21 \\
    & NO  & -0.02 & 0.30 & -1.2$\times10^{-3}$ & -4.7$\times10^{-4}$ & -3.1$\times10^{-4}$ & 1.2$\times10^{-5}$ & 3.4$\times10^{-4}$ & 5.1$\times10^{-4}$ & 1.2$\times10^{-3}$ \\
    & NH  & -0.08 & 1.54 & -1.77 & -0.18 & -0.07 & -3.1$\times10^{-7}$ & 0.06 & 0.12 & 2.11 \\
    & NX  & -0.04 & 0.84 & -0.86 & -0.13 & -0.07 & 6.1$\times10^{-7}$ & 0.07 & 0.11 & 0.53 \\
\midrule
\multicolumn{1}{l}{$q$} & NK  & 6.48 & 13.79 & 3.0$\times10^{-9}$ & 1.1$\times10^{-6}$ & 4.1$\times10^{-6}$ & 4.9$\times10^{-4}$ & 31.24 & 37.10 & 55.80 \\
    & NO  & 1.7$\times10^{-3}$ & 0.03 & 7.2$\times10^{-11}$ & 4.2$\times10^{-10}$ 
& 1.5$\times10^{-9}$ & 2.1$\times10^{-7}$ & 1.4$\times10^{-6}$ & 2.1$\times10^{-6}$ 
& 7.8$\times10^{-5}$ \\
    & NH  & 0.96 & 6.39 & 5.6$\times10^{-9}$ 
& 7.3$\times10^{-5}$ & 2.5$\times10^{-4}$ & 4.2$\times10^{-3}$ & 0.06 & 0.23 & 40.30 \\
    & NX  & 0.49 & 4.97 & 5.0$\times10^{-9}$ & 3.5$\times10^{-6}$ 
& 2.9$\times10^{-4}$ & 0.01 & 0.07 & 0.17 & 3.91 \\
\midrule
\multicolumn{1}{l}{Inflection} & NK  & 28  & 3,136 & -1,615 & -281 & -137 & -0.02 & 123 & 276 & 7,749 \\
Point & NO  & -1,654 & 40,868 & -137,606 & -75,000 & -3,988 & -46 & 3,919 & 43,923 & 132,584 \\
(contracts)    & NH  & -3  & 267 & -199 & -42 & -19 & 0.03 & 21  & 48  & 323 \\
    & NX  & -3  & 272 & -961 & -39 & -17 & -0.01 & 17  & 42  & 792 \\
\midrule
\multicolumn{1}{l}{Adjusted} & NK  & 15.71 & 8.12 & 1.72 & 3.77 & 5.89 & 15.27 & 26.39 & 29.18 & 35.05 \\
\multicolumn{1}{l}{ $R^2$} & NO  & 28.80 & 8.57 & 10.07 & 14.22 & 17.19 & 29.44 & 39.15 & 41.29 & 47.32 \\
\multicolumn{1}{l}{(\%)} & NH  & 15.19 & 5.67 & 3.77 & 6.68 & 8.36 & 14.85 & 22.61 & 25.03 & 28.69 \\
    & NX  & 17.71 & 5.70 & 5.68 & 8.68 & 10.52 & 17.44 & 25.22 & 27.20 & 32.06 \\
\bottomrule
\end{tabular}
\end{footnotesize}
\end{table}
\noindent
{\small
The parameter estimates of the S-shape function are obtained by a constrained
nonlinear least squares procedure on 1-minute data, 
which are processed from Bloomberg's tick data.
The sample period is from early November 2013 to early June 2017,
for a total of 885 days for JPX's onshore futures (NK and NO) 
and 927 days for CME's offshore futures (NH and NX).
A set of parameter estimates is obtained for each trading day in the sample period,
along with the inflection point and the adjusted $R^2$.
We report the daily average, standard deviation (STD), and the percentiles.
}

\clearpage

\begin{table}[htbp]
\centering
\caption{Descriptive Statistics of Best Bid and Ask Sizes}
\vskip1em
\label{t:dsbest}
\begin{center}
\begin{small}
\begin{tabular}{rlrrrrrrrrr}
\toprule
 &  &  &  & \multicolumn{7}{c}{Percentile} \\
\cmidrule{5-11}  &   & \multicolumn{1}{c}{\textbf{Average}} & \multicolumn{1}{c}{\textbf{STD}} & \multicolumn{1}{c}{\textbf{1}} & \multicolumn{1}{c}{\textbf{5}} & \multicolumn{1}{c}{\textbf{10}} & \multicolumn{1}{c}{\textbf{50}} & \multicolumn{1}{c}{\textbf{90}} & \multicolumn{1}{c}{\textbf{95}} & \multicolumn{1}{c}{\textbf{99}} \\
\midrule
\multicolumn{1}{l}{NK} 
& Bid Size & 125  & 44   & 49   & 64   & 73   & 119  & 184  & 208  & 252 \\
& Ask Size & 127  & 45   & 51   & 63   & 73   & 122  & 186  & 209  & 266 \\
\midrule
\multicolumn{1}{l}{NO} 
& Bid Size & 332  & 119  & 111  & 160  & 193  & 327  & 477  & 559  & 668 \\
& Ask Size & 341  & 128  & 110  & 165  & 193  & 330  & 506  & 575  & 720 \\
\midrule
\multicolumn{1}{l}{NH} 
& Bid Size & 8    & 3    & 3    & 4    & 5    & 7    & 12   & 14   & 18 \\
& Ask Size & 8    & 3    & 2    & 4    & 4    & 8    & 12   & 14   & 18 \\
\midrule
\multicolumn{1}{l}{NX} 
& Bid Size & 7    & 3    & 2    & 3    & 4    & 6    & 10   & 12   & 15 \\
& Ask Size & 7    & 3    & 2    & 3    & 4    & 7    & 10   & 13   & 15 \\
\bottomrule
\end{tabular}%
\end{small}
\end{center}
\end{table}
\noindent
{\small
From Bloomberg's tick data, we identify the best bid and ask sizes
at the beginning of each 1-minute interval for each trading session
from 9 AM to 3 PM Japan Standard Time.
The sample period is from early November 2013 to early June 2017,
for a total of 885 days for JPX's onshore futures (NK and NO) 
and 927 days for CME's offshore futures (NH and NX).
Daily average bid and ask sizes are computed
and the table presents the descriptive statistics for the sample period
with ``std'' denoting the standard deviation. 
}


\clearpage

\clearpage

\begin{figure*}[t]
\centering
\vfill
{\small Panel C: CME's Nikkei 225 Yen-Denominated Contract (NH)}

\includegraphics[height=0.44\textheight]{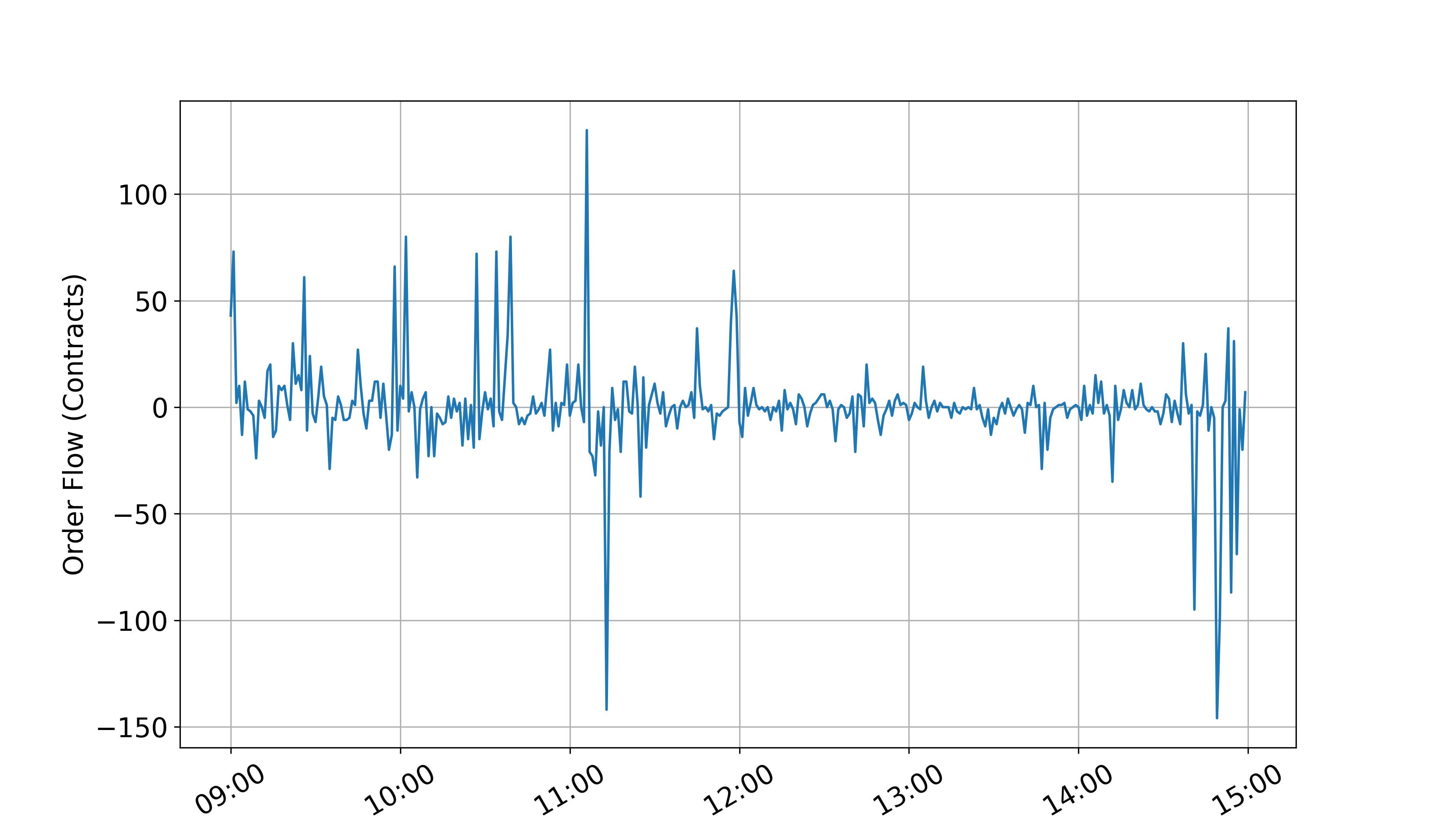}

\bigskip\bigskip

\centering

{\small Panel D: CME's Nikkei 225 Dollar-Denominated Contract (NX)}

\includegraphics[height=0.44\textheight]{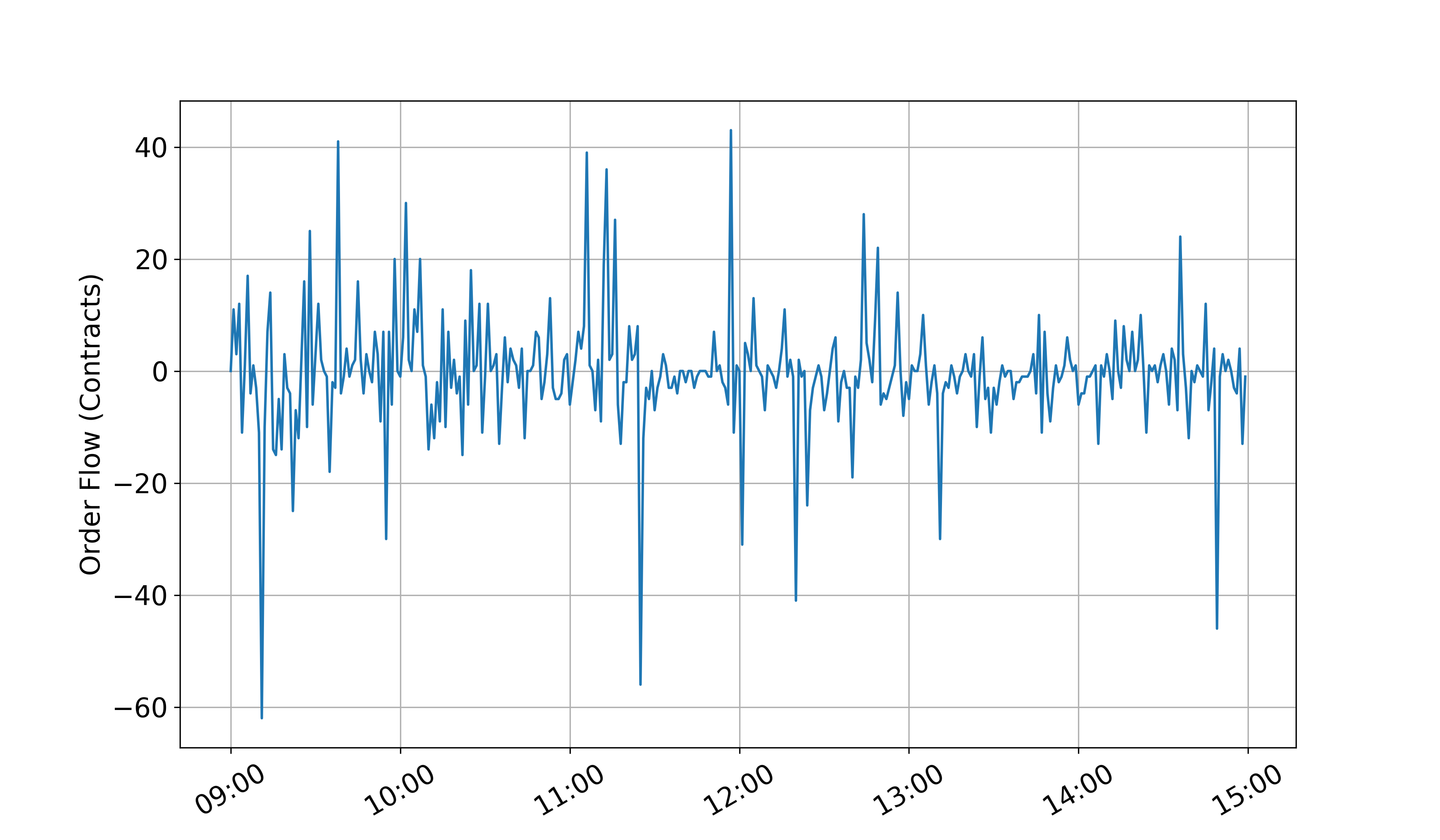}
\vfill
\end{figure*}

\clearpage

\vfil
\clearpage

\vfil
\end{document}